\documentclass[12pt]{article}
\usepackage{cite}

\usepackage{amssymb,latexsym}

\usepackage{epstopdf}

\usepackage{graphics,epsfig}

\usepackage{color}

\usepackage{amsmath,amsfonts}

\usepackage{mathrsfs}

\DeclareMathAlphabet{\mathpzc}{OT1}{pzc}{m}{it}

\usepackage{tikz}
\usetikzlibrary{arrows,shapes}
\usetikzlibrary{trees}
\usetikzlibrary{matrix,arrows} 				% For commutative diagram
\usetikzlibrary{positioning}				% For "above of=" commands
\usetikzlibrary{calc,through}				% For coordinates
\usetikzlibrary{decorations.pathreplacing}  % For curly braces
\usepackage{pgffor}							% For repeating patterns

\usetikzlibrary{decorations.pathmorphing}	% For Feynman Diagrams
\usetikzlibrary{decorations.markings}
\tikzset{
    vector/.style={decorate, decoration={snake}, draw},
	provector/.style={decorate, decoration={snake,amplitude=2.5pt}, draw},
	antivector/.style={decorate, decoration={snake,amplitude=-2.5pt}, draw},
    fermion/.style={draw=black, postaction={decorate},
        decoration={markings,mark=at position .55 with {\arrow[draw=black]{>}}}},
    fermionbar/.style={draw=black, postaction={decorate},
        decoration={markings,mark=at position .55 with {\arrow[draw=black]{<}}}},
    fermionnoarrow/.style={draw=black},
    gluon/.style={decorate, draw=black,
        decoration={coil,amplitude=4pt, segment length=5pt}},
    scalar/.style={dashed,draw=black, postaction={decorate},
        decoration={markings,mark=at position .55 with {\arrow[draw=black]{>}}}},
    scalarbar/.style={dashed,draw=black, postaction={decorate},
        decoration={markings,mark=at position .55 with {\arrow[draw=black]{<}}}},
    scalarnoarrow/.style={dashed,draw=black},
    electron/.style={draw=black, postaction={decorate},
        decoration={markings,mark=at position .55 with {\arrow[draw=black]{>}}}},
	bigvector/.style={decorate, decoration={snake,amplitude=4pt}, draw},
}

\tikzstyle{block} = [draw, rectangle, 
    minimum height=3em, minimum width=6em]

\usepackage{pgfplots}

\usetikzlibrary{positioning,arrows,patterns}
\usetikzlibrary{decorations.markings}
\usetikzlibrary{calc} 

\tikzset{
  photon/.style={decorate, decoration={snake}, draw=black},
  fermion/.style={draw=black, postaction={decorate},decoration={markings,mark=at position .55 with {\arrow{>}}}},
  vertex/.style={draw,shape=circle,fill=black,minimum size=3pt,inner sep=0pt},
}

\usepackage{feynmp}
\usepackage{feynmp-auto}

\newcommand*{\Scale}[2][4]{\scalebox{#1}{$#2$}}%

\let\a=\alpha \let\b=\beta \let\g=\gamma \let\d=\delta \let\e=\epsilon
\let\z=\zeta \let\h=\eta \let\th=\theta  \let\k=\kappa
\let\l=\lambda \let\m=\mu \let\n=\nu \let\x=\xi \let\p=\pi %\let\r=\rho
\let\s=\sigma   \let\f=\phi  
 
\let\w=\omega      \let\G=\Gamma \let\D=\Delta \let\Th=\Theta \let\L=\Lambda
\let\X=\Xi  \let\S=\Sigma  \let\Y=\Psi
 
\let\la=\label  
  
\def\nn{\nonumber} \def\bd{\begin{document}} \def\ed{\end{document}}
\def\ds{\documentstyle} \let\fr=\frac \let\bl=\bigl \let\br=\bigr
\let\Br=\Bigr \let\Bl=\Bigl
\let\bm=\bibitem
\let\na=\nabla
\def\tU{{\widetilde U}}
\let\pa=\partial \let\ov=\overline
\def\ie{{\it i.e.\ }}
\newcommand{\be}{\begin{equation}}
\newcommand{\ee}{\end{equation}}
\def\ba{\begin{array}}
\def\ea{\end{array}}
\def\ft#1#2{{\textstyle{{\scriptstyle #1}\over {\scriptstyle #2}}}}
\def\fft#1#2{{#1 \over #2}}
\def\F#1#2{{ F_{#1}^{(#2)} }}
\def\cF#1#2{{ {\cal F}_{#1}^{(#2)} }}

\def\R{{\bf R}}
\def\sst#1{{\scriptscriptstyle #1}}
\def\oneone{\rlap 1\mkern4mu{\rm l}}
\def\e7{E_{7(+7)}}
\def\td{\tilde}
\def\wtd{\widetilde}
\def\im{{\rm i}}
\def\bog{Bogomol'nyi\ }
\newcommand{\ho}[1]{$\, ^{#1}$}
\newcommand{\hoch}[1]{$\, ^{#1}$}
\newcommand{\bea}{\begin{eqnarray}}
\newcommand{\eea}{\end{eqnarray}}
\newcommand{\ra}{\rightarrow}
\newcommand{\lra}{\longrightarrow}
\newcommand{\Lra}{\Leftrightarrow}
\newcommand{\ap}{\alpha^\prime}
\newcommand{\bp}{\tilde \beta^\prime}
\newcommand{\cB}{{\cal B}}
\newcommand{\cO}{{\cal O}}
\newcommand{\vecx}{\vec{x}}
\newcommand{\vecy}{\vec{y}}
\newcommand{\vecp}{\vec{p}}
\newcommand{\vecq}{\vec{q}}
\newcommand{\tr}{{\rm tr} }
\newcommand{\Tr}{{\rm Tr} }
\newcommand{\NP}{Nucl. Phys. }

\newcommand{\cL}{{\cal L}}
\newcommand{\cA}{{\cal A}}
\newcommand{\cT}{{\cal T}}
\newcommand{\cR}{{\cal R}}
\newcommand{\cD}{{\cal D}}
\newcommand{\cH}{{\cal H}}

\def\Cb{\bar{C}}

\def\sst#1{{\scriptscriptstyle #1}}
\def\0{{\sst{(0)}}}
\def\1{{\sst{(1)}}}
\def\2{{\sst{(2)}}}
\def\3{{\sst{(3)}}}
\def\4{{\sst{(4)}}}
\def\5{{\sst{(5)}}}
\def\6{{\sst{(6)}}}
\def\7{{\sst{(7)}}}
\def\8{{\sst{(8)}}}
\def\9{{\sst{(9)}}}
\def\p{{\sst{(p)}}}
\def\q{{\sst{(q)}}}
\def\ve{\varepsilon}
\def\vf{\varphi}
\def\F{\Phi}
\def\wg{\wedge}

\def\thb{\bar{\theta}}
\def\Thb{\bar{\Theta}}
\def\barp{\bar{p}}
\def\barq{\bar{q}}
\def\barc{\bar{c}}
\def\bard{\bar{d}}
\def\e{\epsilon}

\def \bi{\bibitem}
\def \la {\label}

\def \l {\lambda}
\def\foot{\footnote}
\def \tl  {{\tilde \l}}
\def \sql {{\sqrt \l}}
\def \adss {$AdS_5 \times S^5$\ }
\newcommand{\rf}[1]{(\ref{#1})}
\def \ov {\over}

\def\th{\theta}
\def\Th{\Theta}
\def\vth{\vartheta}
\def\btheta{{\bar\theta}}
\def\ttheta{{{\tilde\theta}}}
\def\bttheta{{{\bar\ttheta}}}
\def\vth{\vartheta}

\def\ra{\rightarrow}
\def\N{\nabla}
\def\F{{\cal F}}
\def\uM{\underline{M}}
\def\uA{\underline{A}}
\def\uN{\underline{N}}
\def\uP{\underline{P}}
\def\ua{\underline{a}}
\def\ub{\underline{b}}
\def\uc{\underline{c}}
\def\ud{\underline{d}}
\def\ue{\underline{e}}
\def\uf{\underline{f}}
\def\ui{\underline{i}}
\def\uj{\underline{j}}
\def\uk{\underline{k}}
\def\ul{\underline{l}}
\def\ual{\underline{\alpha}}
\def\ube{\underline{\beta}}
\def\um{\underline{m}}
\def\un{\underline{n}}
\def\up{\underline{p}}
\def\uq{\underline{q}}
\def\ur{\underline{r}}
\def\us{\underline{s}}
\def\umu{\underline{\mu}}
\def\unu{\underline{\nu}}
\def\ula{\underline{\l}}
\def\uka{\underline{\k}}
\def\usi{\underline{\s}}
\def\urh{\underline{\r}}
\def\cc{\circ}
\def\eqv{\equiv}

\def\ni{\noindent}

\def\Ep{E^{{}^{(+)}}}
\def\Em{E^{{}^{(-)}}}

\def\Mp{M^{{}^{(+)}}}
\def\Mm{M^{{}^{(-)}}}

\def \ha{{1\ov 2}}

\def\r{\rho}

\def\Y{{\rm Y}}
\def\X{{\rm X}}
\def\tY{\tilde{\rm Y}}
\def\tX{\tilde{\rm X}}
\def\dY{\dot{\rm Y}}
\def\dX{\dot{\rm X}}

\def \J {\mathcal{J}}
\def \del {\partial}

\def\dF{\dot{F}}
\def\dG{\dot{G}}
\def\df{\dot{f}}
\def \E {{\cal E}}
\def \S {{\cal S}}
\def \J {{\cal J}}

\def\ms{\mathcal{S}}
\def\mj{\mathcal{J}}
\def\soj{\fr{\ms}{\mj}}
\def \R {{\bf R}}
\def \om {\omega}
\def \bE {\bar E}
\def \x {{\cal X}}

\def \bi{\bibitem}
\def \la {\label}

\def \l {\lambda}
\def\foot{\footnote}
\def \tl  {{\tilde \l}}
\def \sql {{\sqrt \l}}
\def \adss {$AdS_5 \times S^5$\ }
\def \ov {\over}

\def \varpi {{\rm w}}

\def\thb{\bar{\theta}}
\def\Thb{\bar{\Theta}}
\def\mb{\bar{\m}}
\def\ab{\bar{\a}}
\def\zb{\bar{z}}
\def\psib{\bar{\psi}}
\def\barp{\bar{p}}
\def\barq{\bar{q}}
\def\barc{\bar{c}}
\def\bard{\bar{d}}
\def\e{\epsilon}
\def\wb{\bar{w}}
\def\lb{\bar{\l}}
\def\Jb{\bar{J}}
\def\Nb{\bar{N}}
\def\Zb{\bar{Z}}
\def\pab{\bar{\pa}}

\def\At{\tilde{A}}
\def\Bt{\tilde{B}}
\def\Ct{\tilde{C}}
\def\Dt{\tilde{D}}
\def\Et{\tilde{E}}
\def\Ft{\tilde{F}}
\def\Gt{\tilde{G}}
\def\Ht{\tilde{H}}
\def\Kt{\tilde{K}}
\def\Mt{\tilde{M}}
\def\Nt{\tilde{N}}
\def\Rt{\tilde{R}}
\def\at{\tilde{a}}
\def\bt{\tilde{b}}
\def\ct{\tilde{c}}
\def\dt{\tilde{d}}
\def\et{\tilde{e}}
\def\ft{\tilde{f}}
\def \ztt{\tilde{\z}}
\def \zetat{\tilde{\zeta}}
\def\htil{\tilde{h}}
\def\gt{\tilde{g}}
\def\nt{\tilde{n}}
\def\mut{\tilde{\mu}}
\def\nut{\tilde{\nu}}
\def\pht{\tilde{\f}}
\def\Phit{\tilde{\Phi}}
\def\vft{\tilde{\vf}}

\def\rht{\tilde{\rho}}

\def\asth{\hat{*}}
\def\phh{\hat{\phi}}

\def\bA{{\bf A}}

\def\ola{\overleftarrow}
\def\ora{\overrightarrow}
\def\alt{\tilde{\a}}

\def\eh{\hat{e}}
\def\eph{\hat{\e}}
\def\ph{\hat{p}}
\def\alh{\hat{\a}}
\def\beh{\hat{\b}}
\def\gah{\hat{\g}}
\def\Fh{\hat{F}}
\def\muh{\hat{\m}}
\def\nuh{\hat{\n}}
\def\thh{\hat{\th}}
\def\rhh{\hat{\r}}
\def\dh{\hat{d}}
\def\ih{\hat{i}}
\def\jh{\hat{j}}
\def\hh{\hat{h}}
\def\nh{\hat{n}}
\def\gh{\hat{g}}
\def\kh{\hat{k}}
\def\deh{\hat{\d}}
\def\wh{\hat{w}}
\def\lah{\hat{\l}}
\def\Ah{\hat{A}}
\def\Gh{\hat{G}}
\def\Kh{\hat{K}}
\def\Nh{\hat{N}}
\def\Rh{\hat{R}}
\def\Ch{\hat{C}}
\def\Omh{\hat{\Omega}}

\def\xh{\hat{x}}

\def\ps{\rlap{\, /}\;\,p }
\def\ks{\rlap{\, /}\;\,k }

\def\gym{g_{YM}}

\def\adot{\dot{a}}
\def\bdot{\dot{b}}
\def\bpa{\bar{\pa}}

\def\pr{\prime}
\def\ssk{\medskip}
\def\clb{\color{blue}}
\def\clr{\color{red}}
\def\clg{\color{green}}
\def\clp{\color{purple}}
\def\bfA{{\bf A}}
\def\bfB{{\bf B}}
\def\bfK{{\bf K}}
\def\bfU{{\bf U}}
\def\bfX{{\bf X}}
\def\bfY{{\bf Y}}
\def\bfZ{{\bf Z}}
\def\bfg{{\bf g}}
\def\bfn{{\bf n}}

\def\bsk{\bigskip}
\def\ssk{\medskip}

\def\Ec{{\cal E}}

\begin{document}

\overfullrule=0pt
\parskip=2pt
\parindent=12pt
\headheight=0in \headsep=0in \topmargin=0in
\oddsidemargin=0in

\vspace{ -3cm}
\thispagestyle{empty}
%\vspace{1cm}
%\begin{flushright}
%Preprint DFPD 01/TH/\\
%hep-th/
%\end{flushright}

 \vspace{0.1cm}

\setcounter{equation}{0}
\setcounter{footnote}{0}
\setcounter{section}{0}

\begin{center}

{\Large\bf Revisit of renormalization of Einstein-Maxwell theory at one-loop}

\vskip 0.8cm

%
%\vspace{0.5cm}
%
%A. J. Nurmagambetov$\,^{\spadesuit}$\let\thefootnote\relax\footnotetext{$^{\spadesuit}$ Also at {\it Karazin Kharkov National University, 4 Svobody Sq., Kharkov, UA 61022} \& {\it Usikov Institute for Radiophysics and Electronics, 12 Proskura St., Kharkov, UA 61085}. } 
%and 

I. Y. Park

%\\
%
%\vspace{0.3cm}
%
%$^{\spadesuit}$
%{\it Akhiezer Institute for Theoretical Physics of
%NSC KIPT,\\
%1 Akademicheskaya St., Kharkov, \\ UA 61108 Ukraine \\
%ajn@kipt.kharkov.ua
%}
%
\vspace{0.3cm}
{\it {}{$^\dagger$}Department of Applied Mathematics,
Philander Smith College %\footnote{Home institute}
                               \\
Little Rock, AR 72223, USA \\
inyongpark05@gmail.com
}

 \vspace{.5cm}

\end{center}

 \vspace{0.1cm}

\begin{abstract}

In a series of recent works based on foliation-based quantization in which renormalizability has been achieved for the physical sector of the theory, we have shown that the use of the standard graviton propagator interferes, due to the presence of the trace mode, with the 4D covariance. A subtlety in the background field method also requires careful handling. This status of the matter motivated us to revisit an Einstein-scalar system in one of the sequels. Continuing the endeavors, we revisit the one-loop renormalization of an Einstein-Maxwell system in the present work. The systematic renormalization of the cosmological and Newton's constants is carried out by applying the refined background field method. 
One-loop beta function of the vector coupling constant is explicitly computed and compared with the literature. 
The longstanding problem of gauge choice-dependence of the effective action is addressed and the manner in which the gauge-choice independence is restored in the present framework is discussed. The formalism also sheds light on background independent analysis. The renormalization involves a metric field  redefinition originally introduced by `t Hooft; with the field redefinition the theory should be predictive.

\end{abstract}
\newpage

%%%%%%%%%%%%%%%%%

%\vspace{.3in}

%\ni {\bf Acknowledgments}

%\ni The research of this work was funded in part by Hangyang University, South Korea.

\section{Introduction}

 A gravitational system (see, e.g., \cite{Birrell,Barvinsky:1985an,Buchbinder,Frolov,Donoghue:1994dn,Mukhanov} for reviews) is much subtler and more complex than a non-gravitational one in many ways. This aspect is manifest in various forms, most notably in the challenges in quantization, which in turn have been spawning various obstructions. 
One can easily name several areas in which a firmer grasp of the quantization would better position one for a more complete treatment. Any study, in particular, the study of black hole information, in which the back reaction of the metric plays (or is expected to play)  an important role, should be an example. The cosmological constant problem is also likely to benefit since it is the vacuum energy the complete understanding of which must be accompanied by handling of its quantum shift. 
Much of the difficulty in the quantization must be attributed to the large amount of gauge symmetry, the diffeomorphism. Therefore, one can reasonably expect that the key to the puzzle should lie largely in proper handling of the gauge symmetry. It has recently been realized that the diffeomorphism symmetry can be tamed, as depicted in Fig. 1, in a manner that accomplishes the renormalizability of gravity in its physical sector\cite{Park:2014tia}; see \cite{Park:2019amz} for a review. The renormalization procedures of pure Einstein gravity and an Einstein-scalar system have been carried out in the new quantization method in \cite{Park:2014noa,Park:2015ota,Park:2015xoa} and \cite{Park:2015ybl,Park:2016zgt}, respectively. We extend and expand those analyses to an Einstein-Maxwell system in this work. 

\begin{figure}
	\hspace{-.3in}
	\centerline{
		\begin{minipage}[b]{10cm}
			\epsfxsize=12cm
			\epsfbox{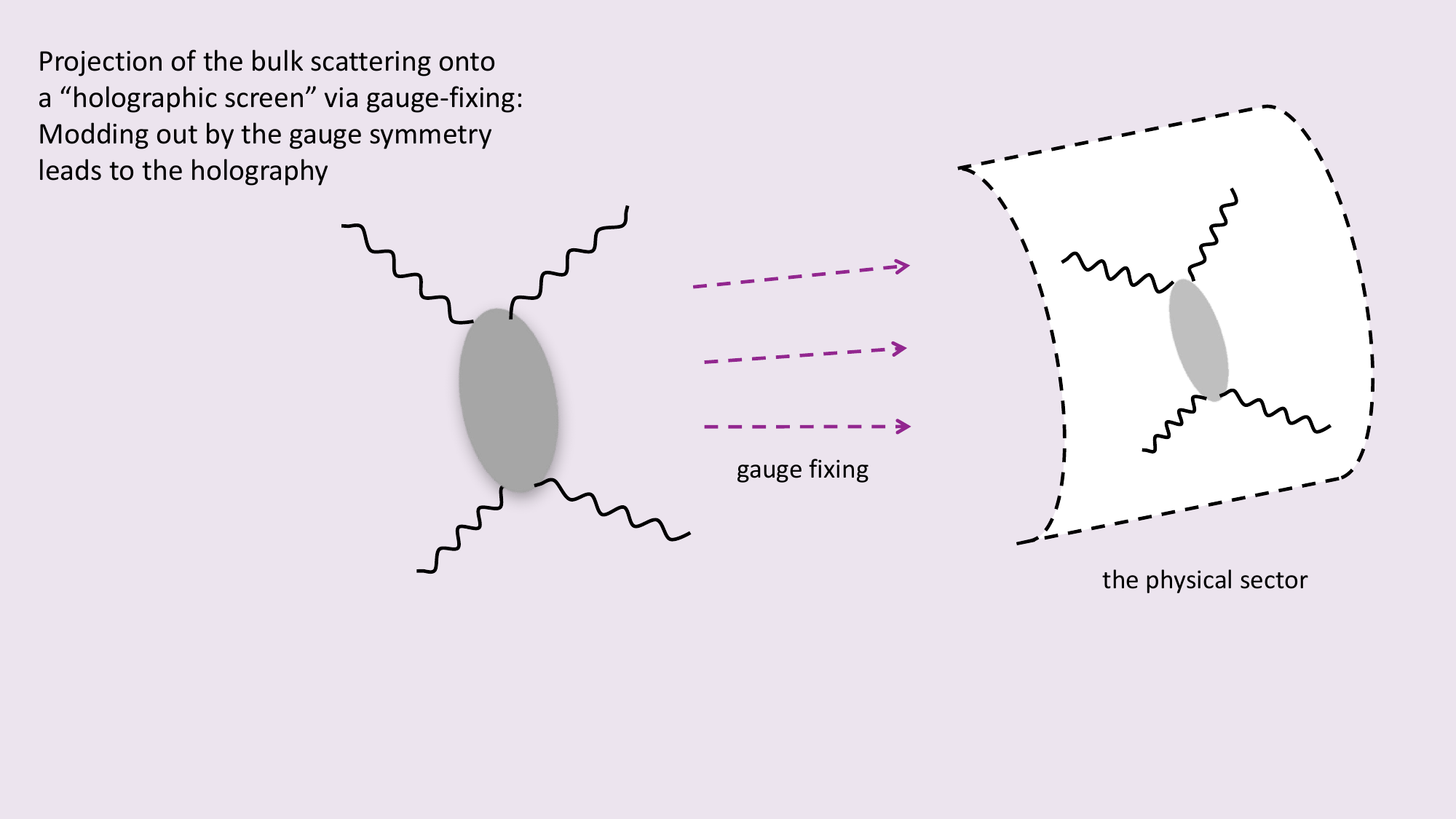}
		\end{minipage}
	}
	\caption{gauge-fixing induced projection onto the physical states}
	\label{fig}
\end{figure}

The difficulties in a gravitational system could foster great opportunity, as, for instance, in holography, for understanding Nature. As is often the case (although nevertheless surprising if true), all these different aspects may not be unrelated and may well in fact hinge closely on one another. Our recent works on the gravity quantization were motivated by the black hole information. While working on the quantization, we have come to realize that our understanding of the boundary conditions and dynamics is as yet incomplete; a more systematic and sound analysis of the boundary conditions needs to precede \cite{Park:2013iqa,Park:2016fxc,Park:2016vam,Nurmagambetov:2018het} a complete treatment of the quantization. As far as we are aware, its seriousness and importance have not, up until recently, been accordingly stressed. (See the recent work by Witten, \cite{Witten:2018lgb}, for a discussion of the boundary conditions.) We have raised the possibility that information may be bleached through a quantum gravitational process in the vicinity of the horizon and released before the entry of the matter into the horizon\cite{Park:2013rm}\cite{Park:2017wiw}. The cosmological constant is generically generated by the loop effects \cite{Park:2016zgt}, as will be reviewed below, and contributes to the generation of time-dependent solutions that in turn are linked with the black hole information \cite{Nurmagambetov:2018het}.

The divergence analysis of an Einstein-Maxwell system was carried out long ago in an extensive work by Deser and van Nieuwenhuizen \cite{Deser:1974cz}. The counter-terms to the ultraviolet divergences were determined essentially by dimensional analysis and covariance. (The precise meaning of this statement will be spelled out in footnote \ref{fnt9} by taking the simpler case considered in \cite{'tHooft:1974bx}.) In our approach they are directly calculated in the Feynman diagrammatic method by employing the ``{\em traceless}" propagator, as opposed to the widely used traceful propagator. {(The need for a traceless propagator was noted in the past \cite{Ogi,Kuchar:1970mu,Gibbons:1978ac,Mazur:1989by,Grillo:1999xp,Ortin}.)} This turns out to be crucial to avoid certain pathologies associated with the trace mode of the fluctuation metric: {as explicitly demonstrated in \cite{Park:2015ota} and \cite{Park:2016zgt}, use of the standard graviton propagator interferes with the 4D covariance.} The results obtained complement the work of \cite{Deser:1974cz} in several aspects. We will also see, as a byproduct, how the long-known gauge dependence issue related to the one noted in \cite{Vilkovisky:1984st,Fradkin:1983nw,Modesto:2017hzl,Huggins:1987zw,Toms:2008dq,Odintsov:1989gz,Odintsov:1991fk,Falls:2015qga} arises and is cleared up (at least) in the present framework.\footnote{We deal with several different facets of the gauge-choice dependence. One of the manifestations is well known, and has been extensively studied in the literature. Another is newly noted in the present framework. They will all be addressed in the body.}

The case analysis of an Einstein-Maxwell system carries several more imminent significances for our perspective. Firstly, the matter part itself is a gauge system and this poses additional hurdles (mentioned, e.g., in footnote \ref{ftn10}); overcoming them should constitute meaningful progress in the field.
Secondly, it is in this work where the field redefinition-utilized renormalization program is more thoroughly carried out: the focus of \cite{Park:2016zgt} was on establishing the {\em renormalizability} itself of a gravity-matter system. A more detailed and explicit analysis of, e.g., the running of the coupling constants was not conducted. In this work, the running of the cosmological constant and Newton's constant is addressed in much detail. Also, the one-loop beta function of the vector coupling constant is explicitly computed to demonstrate the powerfulness of the method. Since the renormalization involves a field redefinition, which is not necessary in the usual renormalizable theories, the explicit steps of the renormalization will be worth presenting - all of the required steps are taken in the present work. Further, the predictability of the theory - brought along by the renormalizability - is also explicitly addressed.

\vspace{.3in}
The paper is organized as follows. 

\vspace{.1in}

\ni In section 2, we outline the one-loop renormalization procedure in a general background metric $g_{\m\n}$ of the metric field equation. (Note that $g_{\m\n}$ denotes a solution, as opposed to the fluctuation denoted by $h_{\m\n}$ and background field by $\vf_{\m\n}$.) The analysis should make it clear that the methodology can be applied to an arbitrary solution  $g_{\m\n}$. The first several relatively simple diagrams and their relevant vertices are identified. In section 3.1, we carry out the explicit one-loop counter-term computation by taking $g_{\m\n}=\h_{\m\n}$.  
A certain diagram yields a non-covariant expression and its inspection leads to a connection with the problem of certain novel gauge choice-dependence\footnote{{This particular gauge-choice dependence issue is new and not the same as the long-known one \cite{Odintsov:1989gz,Odintsov:1991fk,Modesto:2017hzl} although they should have a similar origin. The latter dependence disappears by going onshell \cite{Falls:2015qga}. The present framework elaborates on the onshell disappearance; see section 3.2.}} of the effective action. This particular gauge choice-dependence is then resolved. The origin of the gauge choice-dependence is found in section 3.2 in the limitation of the background field method (BFM) in general, which can alternatively be viewed as a reflection of the complexity of a gravitational system. {Afterwards, we look into how the well-known gauge-choice dependence can be avoided as well. We end the section by noting that the freedom in choosing renormalization conditions is of great aid for facilitating background-independent analysis.}
In section 4 we consider renormalization of the cosmological, Newton's, and vector coupling constants. The vacuum-to-vacuum and tadpole diagrams are responsible for their renormalization. Unlike in a non-gravitational theory, the tadpole diagrams play a potentially important role. There are several technical subtleties, some of which have to do with dimensional regularization: the flat propagator yields vanishing results for the vacuum-to-vacuum and tadpole diagrams. The shifts in the coupling constants are introduced through finite renormalization.  We show that the original Einstein-Hilbert action with the counter-terms can be rewritten as the same form of the Einstein-Hilbert action but now in terms of a redefined metric. An analysis of renormalization of the matter coupling was carried out in \cite{Huggins:1987zw} by employing the setup of \cite{Vilkovisky:1984st}. In the present work the beta function of the matter coupling is carried out by taking the cosmological constant as the graviton mass term in the manner to be explained in the main body. The analysis yields the same result as that of \cite{Toms:2008dq}. Several ramifications including the theory's predictability are discussed. Section 5 contains a summary and future directions. We contemplate on the several possible procedures of renormalization. We also comment on the higher-loop extension of the present work.

%%%%%%%%%%%%%%%%%%%%%%%%%%%%%%%%%%%%%%%%%%%%
%%%%%%%%%%%%%%%%%%%%%%%%%%%%%%%%%%%%%%%%%%%%
\section{Setup of loop computation}
%%%%%%%%%%%%%%%%%%%%%%%%%%%%%%%%%%%%%%%%%%%%
%%%%%%%%%%%%%%%%%%%%%%%%%%%%%%%%%%%%%%%%%%%%

The preliminary step for renormalization and beta functions is to compute the one-particle-irreducible (1PI) effective action in the given background. (See, e.g., \cite{Kallosh:1978wt}\cite{Capper:1984qq}\cite{Buchbinder}\cite{Antoniadis:1995fc} for reviews of various methods of computing the effective action.) In this section, we lay broader outlines of the counter-term computation in an arbitrary background, i.e., a solution of the metric field equation $g_{\m\n}$ before getting into the flat case in section 3.1. (We will come back to the arbitrary metric case in section \ref{sogbi}.) We focus on several two-point amplitudes.

Let us consider the Einstein-Maxwell action,\footnote{To carry out renormalization, one starts with the renormalized form of the action:
\bea
S=\int \sqrt{-\gh_r}\;\Big(\fr1{\k_r^2} \Rh_r-\fr14 \Fh_{r\m\n}^2 \Big)  \la{EM}
\eea
where the renormalized quantities are indicated by the subscript $r$ that has been omitted in \rf{EM2} for simplicity of notation.}
\bea
S=\int \sqrt{-\gh}\;\Big(\fr1{\k^2}\Rh-\fr14 \Fh_{\m\n}^2 \Big). \la{EM2}
\eea
For the perturbative analysis in the background field method (BFM), introduce the fluctuation fields, $(h_{\m\n}, a_\m)$, according to
\bea
\gh_{\m\n}\equiv  h_{\m\n}+\tilde{g}_{\m\n}\quad,\quad
\Ah_\m \equiv a_\m+\At_\m.
  \la{gshift}
\eea
The graviton propagator associated with the {\em traceless} fluctuation mode \cite{Park:2014tia,Park:2015ota,Park:2015xoa} (see also \cite{Morris:2018axr})) can be written as
%%% 
\bea 
<h_{\m\n}(x_1)h_{\r\s}(x_2)>&=& \tilde{P}_{\m\n\r\s}\, \tilde{\D}(x_1-x_2)  \la{h2pt}
\eea
%%%
where the tensor $\tilde{P}_{\m\n\r\s}$ is given by
%%%
\bea
\tilde{P}_{\m\n\r\s} &\equiv& \fr{(2\k^2)}2\Big(\gt_{\m\r}\gt_{\n\s}+\gt_{\m\s}\gt_{\n\r}
- \fr12\gt_{\m\n}\gt_{\r\s}\Big);   \la{fpt}
\eea
%%%
$\tilde{\D}(x_1-x_2)$ is the Green's function for a scalar theory in the background metric $\gt_{\m\n}$. (There is of course the full propagator for the vector field; we will focus on the graviton sector.) {The propagator almost exclusively used in the literature has a {\em trace piece}: instead of the  coefficient $-1/2$ inside parentheses of \rf{fpt}, the standard propagator has $-1$. As demonstrated in our previous works (e.g., \cite{Park:2014noa,Park:2015ota,Park:2016zgt}), the use of the traceful propagator destroys 4D covariance.} As a matter of fact, the need for a traceless propagator was observed in the past \cite{Ogi,Kuchar:1970mu,Gibbons:1978ac,Mazur:1989by,Grillo:1999xp,Ortin}. Our results explicitly demonstrate the pathology associated with the trace piece in the Feynman diagrammatic computation. 

It turns out convenient to employ two different layers of perturbation.  As we will see, it is possible to formally construct $\tilde{\D}(x_1-x_2)$ in a closed-form; one may compute some of the diagrams by employing the full propagator \rf{h2pt} (as well as the full propagator of the Maxwell sector) - which we call the ``first-layer" perturbation. {(More on this ``one-stroke" method later.)} For other diagrams, in particular pure graviton diagrams, one may employ the ``second-layer" perturbation\footnote{For the one-loop effective action computation the present method correspond to perturbative expansion of the result obtained by applying the usual determinant formula, $\int D\xi \;e^{-\fr12 \xi\, K \,\xi}= e^{-\fr12\tr \ln \fr{K}{2\pi}}$. The second-layer perturbation is not necessary in non-gravitational theories.} by splitting $\tilde{g}, \At_\m$ appearing in the right-hand side of \rf{gshift} into
\bea
\quad \tilde{g}_{\m\n} \equiv \vf_{\m\n}+g_{\m\n}\quad,\quad \At_\m \equiv A_\m+A_{0\m} \la{split}
\eea
where $\vf_{\m\n},A_\m$ represents the background fields and $g_{\m\n}, A_{0\m}$  the classical solutions. (For instance, we will take $g_{\m\n}=\eta_{\m\n}, A_{0\m}=0$ in section 3.) {The shift with \rf{split} has been dubbed `double-shift' in \cite{Park:2015ota}. Essentially the same procedure was employed, though a bit implicitly, in \cite{Kallosh:1978wt,Capper:1984qq,Antoniadis:1995fc}.} The need for the second layer perturbation for the gravity sector was discussed, e.g., in \cite{Park:2014noa} and \cite{Park:2015ota}. For most of the diagrams that we will consider, the structures of the vertices allow one to approximate $\tilde{P}_{\m\n\r\s}$, for the given order
%%%
\bea
\tilde{P}_{\m\n\r\s} \simeq P_{\m\n\r\s} 
\equiv \fr{(2\k^2)}2\Big(g_{\m\r}g_{\n\s}+g_{\m\s}g_{\n\r}
- \fr12g_{\m\n}g_{\r\s}\Big)  \la{apt}
\eea
where $P_{\m\n\r\s}$ is the leading-order $\vf_{\m\n}$-expansion of $\tilde{P}_{\m\n\r\s}$. We will also see the use of the full tensor $\tilde{P}_{\m\n\r\s}$ in some of the computations, the first-layer perturbation examples. For the divergence analysis one can use $\tilde{\D}(x_1-x_2)\simeq \D(x_1-x_2)$ where $ \D(x_1-x_2)$ denotes the scalar propagator for $g_{\m\n}=\h_{\m\n}$, 
\bea
\D(x_1-x_2)=\int \fr{d^4k}{(2\pi)^4}\fr{e^{ik\cdot (x_1-x_2)}}{i k^2}.
\eea
In this ``bottom-up" approach, the quantities that one intends to calculate in the first-layer perturbation can be calculated through the second-layer perturbation. Dimensional analysis and 4D covariance provide useful consistency checks as will be demonstrated in section 3.

\vspace{.1in}

Let us  expand the action in terms of the fluctuation fields $h_{\m\n}, a_\m$.
Including the gauge-fixing and ghost terms, one gets
\bea
S=\int \Big( \fr1{\k^2}\cL_{grav}+\cL_{matter}\Big) \la{combinedaction}
\eea
where\footnote{The $\Rt_{\m\n}\bar{C}^\m C^\n$ term of the gravity sector action $\cL_{grav}$ presented in \cite{Park:2016zgt} has a sign error due to mixed conventions. It, with affected equations, has been corrected in \cite{Park:2015ota}.}
\bea
&&\k^2\cL_{grav} =\fr1{2} \sqrt{-\gt}\,\Big( -\fr12\tilde{\N}_\g h^{\a\b}\tilde{\N}^\g h_{\a\b}+\fr14 \tilde{\N}_\g h^{\a}_\a \tilde{\N}^\g h^{\b}_\b  
\nn\\
&&+h_{\a\b}h_{\g\d}\Rt^{\a\g\b\d}-h_{\a\b}h^{\b}{}_\g \Rt^{\k\a\g}{}_{\k}
{ -}h^{\a}{}_{\a}h_{\b\g}\Rt^{\b\g}-\fr12 h^{\a\b}h_{\a\b}\Rt
+\fr14  h^{\a}_\a  h^{\b}_\b \Rt +\cdots\Big) \nn\\
&&-\tilde{\N}^\n \Cb^\m \tilde{\N}_\n C_\m  { +}\Rt_{\m\n}\bar{C}^\m C^\n   -\w^* \tilde{\N}^\m\Ft_{\m\n}C^\n-\w^* \Ft_{\m\n} \tilde{\N}^\m C^\n
+\cdots 
\eea
and
\bea
&&\hspace{-.3in}\cL_{matter} =-\fr14 \sqrt{-\gt}\Big[\gt^{\m\n}\gt^{\r\s}-\gt^{\m\n}h^{\r\s} -\gt^{\r\s}h^{\m\n}+\fr12 \gt^{\m\n}\gt^{\r\s}h
+\gt^{\m\n}h^{\r\k}h_\k^{\s}  +\gt^{\r\s}h^{\m\k}h_\k^{\n}\nn\\
&&\hspace{-.7in}  -\fr12 \gt^{\m\n}hh^{\r\s}  -\fr12 \gt^{\r\s}hh^{\m\n}+h^{\m\n}h^{\r\s}  
+\fr18 \gt^{\m\n} \gt^{\r\s}(h^2-2h_{\k_1\k_2}h^{\k_1\k_2} )
\Big] \Big( f_{\m\r}f_{\n\s} {+}  2f_{\m\r}\Ft_{\n\s}+\Ft_{\m\r}\Ft_{\n\s} \Big) \nn\\
&&
\hspace{.5in}-\fr12\sqrt{-\gt}\; (\tilde{\N}_\k a^\k)^2 -\tilde{\N}\w^* \tilde{\N}\w+\cdots 
\eea
where the raising and lowering are done by $\gt^{\m\n}$ and $\gt_{\m\n}$, respectively.
Above, $(C^\k, \w)$ are the ghosts for the diffeomorphism and vector gauge transformation, respectively.\footnote{These ghost terms correspond to the following transformations of the fluctuation fields \cite{Deser:1974cz}:
\bea
h'_{\m\n}&=&h_{\m\n} +(\gt_{\m\k} \Dt_\n+\gt_{\n\k} \Dt_\m)\eta^\k
            +(h_{\m\k} \Dt_\n+h_{\n\k} \Dt_\m)\eta^\k +\eta^\k \Dt_\k h^{\m\n} \nn\\
a'_\m&=& a_\m+\eta^\k \Ft_{\k\m}+\Dt_\m \eta^5+a_\k\Dt_\m\eta^\m+\eta^\k \Dt_\k a_\m
\eea
under $x'^{\a}=x^\a-\eta^\a$ and the vector gauge transformation with the parameter $-\eta^\k \At_\k+\eta^5$.
}
Putting it all together,  \rf{combinedaction} can be written in a more useful form as the sum of the kinetic part and the vertices:
\bea
S\equiv S_{k}+S_{v}
\eea
with 
\bea
\hspace{-.5in}S_{k}&\!\!=&\!\! \int \sqrt{-\gt}\,   \fr1{2\k^2}  \Big( -\fr12\tilde{\N}_\g h^{\a\b}\tilde{\N}^\g h_{\a\b}+\fr14 \tilde{\N}_\g h^{\a}_\a \tilde{\N}^\g h^{\b}_\b \Big)   -\fr14 \sqrt{-\gt}\;\Big(\gt^{\m\n}\gt^{\r\s}f_{\m\r}f_{\n\s}\Big)   
                                                    \nn\\
       && -\fr12\sqrt{-\gt}\; (\tilde{\N}_\k a^\k)^2+ \fr1{2\k^2}\sqrt{-\gt}\; (-\tilde{\N}^\n \Cb^\m \tilde{\N}_\n C_\m)-\sqrt{-\gt}\;\tilde{\N}^\r\w^* \tilde{\N}_\r\w            \la{Sk}                                 
\eea
and
\bea                                                    
 S_{v}&=&\int\sqrt{-\gt}\;\fr1{2\k^2}   \Big(h_{\a\b}h_{\g\d}\Rt^{\a\g\b\d}-h_{\a\b}h^{\b}{}_\g \Rt^{\k\a\g}{}_{\k}
{ -}h^{\a}{}_{\a}h_{\b\g}\Rt^{\b\g}-\fr12 h^{\a\b}h_{\a\b}\Rt  \nn\\
&&+\fr14  h^{\a}_\a  h^{\b}_\b \Rt \Big) 
-\fr14 \sqrt{-\gt}(\gt^{\m\n}\gt^{\r\s}) \Big( {+}  2f_{\m\r}\Ft_{\n\s}+\Ft_{\m\r}\Ft_{\n\s} \Big) -\fr14 \sqrt{-\gt}\Big[-\gt^{\m\n}h^{\r\s}\nn\\
&&  -\gt^{\r\s}h^{\m\n}+\fr12 \gt^{\m\n}\gt^{\r\s}h
+\gt^{\m\n}h^{\r\k}h_\k^{\s}  +\gt^{\r\s}h^{\m\k}h_\k^{\n} -\fr12 \gt^{\m\n}hh^{\r\s}  -\fr12 \gt^{\r\s}hh^{\m\n}\nn\\
&&+h^{\m\n}h^{\r\s}  
+\fr18 \gt^{\m\n} \gt^{\r\s}(h^2-2h_{\k_1\k_2}h^{\k_1\k_2} )
\Big] \Big( f_{\m\r}f_{\n\s} {+}  2f_{\m\r}\Ft_{\n\s}+\Ft_{\m\r}\Ft_{\n\s} \Big) \nn\\
&& \hspace{.5in}+ \fr1{2\k^2}\sqrt{-\gt}\; \Bigg(  \Rt_{\m\n}\bar{C}^\m C^\n 
  + \fr12\tilde{\N}^\m\w^* \Ft_{\m\n}C^\n \Bigg)    +\cdots. 
   \la{Sv}
\eea

%%%%%%%%%%%%%%%%%%%%%%%%%%%%%%%%%%%%%%%%%%%%
\subsection{on the gauge-fixing}
%%%%%%%%%%%%%%%%%%%%%%%%%%%%%%%%%%%%%%%%%%%%

%As stated in the introduction, the quantization of an Einstein-Maxwell has additional complications.
A crucial feature of the action above - which has been set up for the refined BFM - is how the graviton gauge-fixing has been implemented:
\bea
-\fr12\Big[\tilde{\nabla}_\n h^{\m\n}-\fr12 \tilde{\nabla}^\m h \Big]^2. \la{bfmgf}
\eea
This is the refined BFM version of the usual gauge-fixing,
\bea
-\fr12\Big[{\nabla}_\n h^{\m\n}-\fr12 {\nabla}^\m h \Big]^2 \la{nbfmgf}
\eea
that is $\gt_{\m\n}$-background non-covariant. In other words, one starts with \rf{nbfmgf} and converts it into \rf{bfmgf} when turning to the refined BFM. The physical content of the gauge condition satisfied by $h_{\m\n}$ is still \rf{nbfmgf} since the BFM is just a convenience device that allows one to conduct the analysis more covariantly than otherwise. (The field $\vf_{\m\n}$ satisfies the same gauge-fixing; see \rf{vfgf} below.) Naively, one expects that with the gauge-fixing \rf{bfmgf} the 1PI effective action will come out to be $\gt_{\m\n}$-covariant. Later we will see that the 1PI action is non-covariant due to the presence of the terms that can be removed by enforcing the strong form of the gauge condition, which provides an important clue as to how to solve {this particular} gauge choice-dependence of the effective action.

%%%%%%%%%%%%%%%%%%%%%%%%%%%%%%%%%%%%%%%%%%%%
\subsection{two-point diagrams}
%%%%%%%%%%%%%%%%%%%%%%%%%%%%%%%%%%%%%%%%%%%%

In general the renormalization in a curved background $g_{\m\n}$ is technically involved. It is nevertheless possible to outline the steps of the amplitude computation for an arbitrary solution metric $g_{\m\n}$.

Cautionary remarks are in order. It is important to distinguish the second-layer diagrams from the first-layer ones. Only the first-layer diagrams will individually yield covariant results. A given first-layer diagram corresponds, in general, to multiple second-layer diagrams even at a fixed order of $\vf_{\m\n}$. 
Consider, for example, the graviton kinetic action,
\bea
\cL_{grav, kin} =\fr1{2\k^2} \sqrt{-\gt}\,\Big( -\fr12\tilde{\N}_\g h^{\a\b}\tilde{\N}^\g h_{\a\b}+\fr14 \tilde{\N}_\g h^{\a}_\a \tilde{\N}^\g h^{\b}_\b  \Big)
\eea
and the one-loop vacuum-to-vacuum amplitude. Although there is a unique one-loop vacuum-to-vacuum amplitude,\!\!\begin{fmffile}{vacandtad2}
\!\!\!\!\Scale[0.4]{
\begin{gathered}
\begin{fmfgraph*}(75,50)\fmfpen{thick}
   \fmfi{gluon}{reverse fullcircle scaled .5w shifted (.5w,.5h)}
  \end{fmfgraph*}\!\!
\end{gathered}},     
\end{fmffile}in the first-layer perturbation, the diagram corresponds to multiple second-layer ones. At the second order in $\vf_{\a\b}$, the relevant diagram is the one given in Fig. 2 (a). More on this as we continue.

With the split given in \rf{split}, the kinetic terms themselves yield the vertices for the second-layer perturbation expansion. For instance, the graviton kinetic term is expanded as  
\bea
 \hspace{.2in}2\k^2 \cL_{grav, kin}= -\fr12 {\pa}_\g h^{\a\b}{\pa}^\g h_{\a\b}+\fr14 {\pa}_\g h^{\a}_\a {\pa}^\g h^{\b}_\b   \la{lv12qq}
\eea
\[
   \hspace{-.2in} + \Big(2g^{\b\b'}\tilde{\G}^{\a' \g\a}- g^{\a\b}\tilde{\G}^{\a' \g\b'}\Big)\pa_\g h_{\a\b}\, h_{\a'\b'}  
 +\Big[\fr12(g^{\a\a'}g^{\b\b'}\vf^{\g\g'}+g^{\b\b'}g^{\g\g'}\vf^{\a\a'} 
 \]
 \[
\hspace{-.2in} +g^{\a\a'}g^{\g\g'}\vf^{\b\b'}) -\fr14 \vf\, g^{\a\a'}g^{\b\b'}g^{\g\g'}-\fr12 g^{\g\g'}g^{\a'\b'}\vf^{\a\b}  
+\fr14 (-\vf^{\g\g'}+\fr12 \vf g^{\g\g'})g^{\a\b}g^{\a'\b'}
\Big] \pa_\g h_{\a\b}\, \pa_{\g'}h_{\a'\b'} 
\]

\ni where the raising and lowering are done by $g^{\m\n}$ and $g_{\m\n}$, respectively. The terms in the second and third lines serve as the vertices responsible for Fig. \ref{fig:2gh} (a). The corresponding ghost diagram is given in Fig. \ref{fig:2gh} (b). 
 \begin{figure}[t]
\begin{center}
\begin{fmffile}{2ghostfig}
\quad\quad 
 \parbox{40mm}{
 \begin{fmfgraph*}(80,50)
\fmfleft{i} \fmfright{o}
\fmf{gluon,tension=6}{i,v1} \fmf{gluon,tension=6}{v2,o}
\fmf{gluon,left,label=$h$,tension=1}{v1,v2,v1}
     \fmflabel{$\vf$}{i}   \fmflabel{$\vf$}{o} \fmf{phantom,label.dist=0}{v1,v2}
\end{fmfgraph*}
}
\quad\quad 
 \parbox{40mm}{
 \begin{fmfgraph*}(80,50)
\fmfleft{i} \fmfright{o}
\fmf{gluon,tension=6}{i,v1} \fmf{gluon,tension=6}{v2,o}
\fmf{dashes,left,label=$C$,tension=1}{v1,v2,v1}
     \fmflabel{$\vf$}{i}   \fmflabel{$\vf$}{o} \fmf{phantom,label.dist=0}{v1,v2}
\end{fmfgraph*}
}
\end{fmffile} 
 \end{center} 
 \vspace{.1in}
          \hspace{1.51in}    (a)    \hspace{1.64in}    (b)
       \caption{\label{fig:2gh}graviton and ghost diagrams (indices on fields suppressed)}
\end{figure}
The forms of all possible second-layer vertices can be obtained by applying this scheme to the rest of the terms in eq. \rf{Sk} and \rf{Sv}. The first several relatively simple matter-involving diagrams are listed in Fig. 2. In general, we restrict the maximum number of the graviton external lines to two for simplicity. Overall, the diagrams are classified into four categories. The first class is the diagrams with both vertices from the graviton sector: the pure gravity sector two-point amplitude and the corresponding ghost-loop diagram in Fig. \ref{fig:2gh}. They were considered in \cite{Park:2015ota} and will be reviewed below.
The second is the diagrams with both vertices from the matter sector, Fig. 3 (a) {to} (c). The third is the diagrams with one vertex from the graviton sector and the other from the matter sector, Fig. 3 (d). 
 
\vspace{.1in}  
\[
%\begin{center}
\begin{fmffile}{13gr2}
\hspace{-.05in} \parbox{40mm}{
 \begin{fmfgraph*}(70,40)
\fmfleft{i} \fmfright{o}
\fmf{gluon,tension=6}{i,v1} \fmf{gluon,tension=6}{v2,o}
\fmf{photon,left,label=$a$,tension=1.3}{v1,v2,v1}
     \fmflabel{$\vf$}{i}   \fmflabel{$\vf$}{o} % \fmf{phantom,label.dist=0,label=$V\;\;\;\;\;V$}{v1,v2}
\end{fmfgraph*}\\
  
          \hspace{.34in}    (a)
}
\parbox{40mm}{
 \begin{fmfgraph*}(70,40)
\fmfleft{i} \fmfright{o}
\fmf{gluon,tension=6}{i,v1} \fmf{gluon,tension=6}{v2,o}
\fmf{dots,left,label=$a$,tension=1.3}{v1,v2,v1}
     \fmflabel{$\vf$}{i}   \fmflabel{$\vf$}{o} %\fmf{phantom,label.dist=0,label=$V\;\;\;\;\;V$}{v1,v2}
\end{fmfgraph*}\\
  
          \hspace{.36in}    (b)
}
\hspace{-.3in}\parbox{40mm}{\begin{fmfgraph*}(78,55)
    \fmfstraight
    \fmfleft{i1,i2} \fmfright{o1,o2}
    \fmfleft{i1,i2}
    \fmfright{o1,o2}
    \fmf{photon}{i1,v1,i2} \fmf{photon}{o1,v2,o2}
    \fmf{gluon,left,label=$h$,tension=.5}{v1,v2,v1}
     \fmflabel{A}{i1}  \fmflabel{A}{i2}  \fmflabel{A}{o1}  \fmflabel{A}{o2}
  \end{fmfgraph*}\\
  
          \hspace{.41in}    (c)
 }  
\parbox{40mm}{\begin{fmfgraph*}(75,55)
    \fmfstraight
    \fmfleft{i1,i2}
    \fmfright{o}
    \fmf{photon}{i1,v1,i2} \fmf{gluon}{v2,o}
    \fmf{gluon,left,label=$h$,tension=.26}{v1,v2,v1}
      \fmflabel{A}{i1}  \fmflabel{A}{i2}  \fmflabel{$\vf$}{o}
  \end{fmfgraph*}\\
  
          \hspace{.3in}    (d)
 }  
\end{fmffile}  
 %\end{center} 
%\end{figure}
\]
\vspace{-.1in}
\[
 \mbox{Figure 3: matter-involving diagrams}
\]
%%%%%%%%%%%%%%%%%%%%%%%%%%
%%%%%%%%%%%%%%%%%%%%%%%%%%
%%%%%%%%%%%%%%%%%%%%%%%%%%
%%%%%%%%%%%%%%%%%%%%%%%%%%
All of the diagrams so far have ``homogeneous" loops whereas the diagrams in Fig. 4 have ``inhomogeneous or heterotic" ones. They are classified as the fourth class due to the fact that they require special care. 
\[
\hspace{-.1in}
%\begin{center}
\parbox{80mm}{
\begin{fmffile}{mixrel}
\Scale[0.99]{
 \begin{fmfgraph*}(90,70)
\fmfleft{i} \fmfright{o}
\fmf{photon,tension=4}{i,v1} \fmf{photon,tension=4}{v2,o}
\fmf{photon,left,tension=1}{v1,v2}  \fmf{gluon,left,tension=1}{v2,v1}
 \fmflabel{$\vf$}{i}   \fmflabel{$\vf$}{o}
\end{fmfgraph*}
}
\quad\quad\quad\quad
\Scale[.9]{
 \begin{fmfgraph*}(90,70)
  \fmfleft{i1,i2,i3,i4}
    \fmfright{o}
 \fmf{phantom,tension=1}{i1,v1}   \fmf{photon,tension=1}{i2,v1}  \fmf{gluon,tension=1}{i3,v1} \fmf{phantom,tension=1}{i4,v1} 
\fmf{photon,tension=3}{v2,o}
\fmf{photon,left,tension=.5}{v1,v2}  \fmf{gluon,left,tension=.8}{v2,v1}
 \fmflabel{$\vf$}{i3} \fmflabel{$\vf$}{o}  \fmflabel{A}{i2}
\end{fmfgraph*}
}
\end{fmffile}
 \vspace{-.2in}
\hspace{0.45in} (a) \hspace{1.69in} (b) \\

\hspace{-.15in}\mbox{ Figure 4: diagrams with inhomogeneous loops}
}
\]
%%%%%%%%%%%%%%%%%%%
\ni The vertex, $V_{g}$, responsible for the diagrams in Fig. \ref{fig:2gh} (a), is defined by rewriting \rf{lv12qq} as
\bea
\cL&=& \fr1{\k'^2}\Big[-\fr12 {\pa}_\g h^{\a\b}{\pa}^\g h_{\a\b}+\fr14 {\pa}_\g h^{\a}_\a {\pa}^\g h^{\b}_\b   +{\cL_{V_{g}}} \Big] \la{eawv}
\eea
where
\bea 
\k'^2\equiv 2\k^2 
\eea
and
\[
\hspace{-.07in} V_{g} \equiv  \sqrt{-g}\Big(2g^{\b\b'}\tilde{\G}^{\a' \g\a}\!-\! g^{\a\b}\tilde{\G}^{\a' \g\b'}\Big)\pa_\g h_{\a\b}\, h_{\a'\b'}  
 \!+\! \sqrt{-g}\Big[\fr12(g^{\a\a'}g^{\b\b'}\vf^{\g\g'}+\!g^{\b\b'}g^{\g\g'}\vf^{\a\a'} 
 \]
 \vspace{-.2in}
 \[
 \hspace{-.3in}
 +g^{\a\a'}g^{\g\g'}\vf^{\b\b'})  -\fr14 \vf\, g^{\a\a'}g^{\b\b'}g^{\g\g'}  -\fr12 g^{\g\g'}g^{\a'\b'}\vf^{\a\b}  
+\fr14 (-\vf^{\g\g'} +\fr12 \vf g^{\g\g'} )g^{\a\b} g^{\a'\b'}
\Big] \pa_\g h_{\a\b}\, \pa_{\g'}h_{\a'\b'} 
\]
 \vspace{-.3in}
\bea
&& \hspace{-.2in}+\sqrt{-\gt}\Big( h_{\a\b}h_{\g\d}\Rt^{\a\g\b\d}-h_{\a\b}h^{\b}{}_\g \Rt^{\k\a\g}{}_{\k}  { -}h^{\a}{}_{\a}h_{\b\g}\Rt^{\b\g}
-\fr12 h^{\a\b}h_{\a\b}\Rt  +\fr14  h^{\a}_\a  h^{\b}_\b \Rt
 \Big).\nn\\
% \vspace{-1in}
\eea
$\cL_{V_{g}}$ appearing in \rf{eawv} and $V_{g}$ are related by $V_{g}=\sqrt{-g}\;\cL_{V_{g}}$. As for $\gt_{\m\n}$-containing quantities, expansion in terms of $\vf_{\m\n}$ is to be understood. 
The vertex responsible for the ghost-loop diagram can be similarly identified by expanding the terms quadratic in the ghost field:
\bea
V_{C} &\equiv &- { \sqrt{-g}} \Big[
\fr{1}{2}\vf \pa^\m \bar{C}^\n \pa_\m {C}_\n
-\tilde{\G}^\l_{\m\n}(\pa^\m \bar{C}^\n { C_\l} -\pa^\m {C}^\n  \bar{C}_\l  ) \nn\\
&&-(g^{\n\b}\vf^{\m\a}+g^{\m\a}\vf^{\n\b})\pa_\b \bar{C}_\a \pa_\n {C}_\m
\Big] {+}{\sqrt{-g}}\; R_{\m\n}\bar{C}^\m C^\n. 
\eea
The vertices responsible for the diagrams in Fig. 3 and Fig. 4 can be similarly obtained by examining the matter part of the action (the trace piece $h\equiv \gt^{\a\b}h_{\a\b}$ has been set to zero \cite{Park:2014tia,Park:2015ota,Park:2015xoa}): 
 \bea
V_{m1} &\equiv&  -\fr14 \sqrt{-g}\Big[ -g^{\r\s}\vf^{\m\n}-g^{\m\n}\vf^{\r\s}
\Big]  f_{\m\r}f_{\n\s}     \nn\\
 V_{m2} &\equiv&  -\fr14 \sqrt{-g}\Big[g^{\m\n}h^{\r\k}h_\k^{\s}  +g^{\r\s}h^{\m\k}h_\k^{\n}
 +h^{\m\n}h^{\r\s}  -\fr14 g^{\m\n} g^{\r\s}h_{\k_1\k_2}h^{\k_1\k_2} 
\Big]  \Ft_{\m\r}\Ft_{\n\s}  \nn\\
 V_{m3} &\equiv& \fr12  \sqrt{-g} \Big[g^{\m\n}h^{\r\s}+g^{\r\s}h^{\m\n}\Big] f_{\m\r}F_{\n\s} \nn\\
 V_{m4}&\equiv& -\fr12 \sqrt{-g} \Big[ \vf^{\m\n}h^{\r\s}+\vf^{\r\s}h^{\m\n}\Big] f_{\m\r}F_{\n\s}.
\eea
%%%%%%%%%%%%%%%%%%%%%%%%%%%
Let us work out the counter-terms to the diagrams in Fig. 2 to 4. Below
\bea
``&\Rightarrow&" \mbox{ means that the diagram on the left-hand side leads to } \nn\\
  && \quad\quad \mbox{the counter-term(s) on the right-hand side}  \nn
\eea
The graviton and ghost contributions respectively are 
{
\bea
\begin{fmffile}{pure1}
\Scale[0.4]{
\begin{gathered}
 \begin{fmfgraph*}(80,50)
\fmfleft{i} \fmfright{o}
\fmf{gluon,tension=6}{i,v1} \fmf{gluon,tension=6}{v2,o}
\fmf{gluon,left,tension=1}{v1,v2,v1}
%
 %    \fmflabel{$\vf$}{i}   \fmflabel{$\vf$}{o} \fmf{phantom,label.dist=0}{v1,v2}
\end{fmfgraph*}
\end{gathered}
}
\end{fmffile}
&\Rightarrow & -\fr12 { \fr{1}{\k'^4}} <\Big(\int V_{g}\Big)^2>
\la{tghpri}
\eea}
and
\bea
\begin{fmffile}{pure2}
\!\!\Scale[0.4]{
\begin{gathered}
 \begin{fmfgraph*}(80,50)
\fmfleft{i} \fmfright{o}
\fmf{gluon,tension=6}{i,v1} \fmf{gluon,tension=6}{v2,o}
\fmf{dashes,left,tension=1}{v1,v2,v1}
\end{fmfgraph*}
\end{gathered}
}
\end{fmffile}  
&\Rightarrow&  -\fr12 { \fr{1}{\k'^4}}<\Big(\int V_{C} \Big)^2>.
\la{tghpri2}
\eea
The numerical {factors, $-\fr12$'s, are the combinatoric factors that arise} when the vertices are brought down from the exponent in the path integral. 
The total gravity sector one-loop counter-terms are given by the sum of these two; the result for the flat case was obtained in \cite{Park:2015ota} will be quoted in section 3.
The diagrams in Fig. 3 (a) and (c) have two vertices, $V_{m1}$ and $V_{m2}$, inserted respectively; one gets
\bea
&&\hspace{-.3in}\begin{fmffile}{res1}
\Scale[0.5]{
\begin{gathered}
\begin{fmfgraph}(80,50)
\fmfleft{i} \fmfright{o}
\fmf{gluon}{i,v1} \fmf{gluon}{v2,o}
\fmf{photon,left,tension=.3}{v1,v2,v1}
\end{fmfgraph}
  \end{gathered} 
} 
\end{fmffile} \Rightarrow -\fr12<\Big(\int V_{m1}\Big)^2>  = -\fr12 <\Big[\int\fr14 ( g^{\r\s}\vf^{\m\n}+g^{\m\n}\vf^{\r\s}) ( f_{\m\r}f_{\n\s}  )\Big]^2> \nn\\
\eea
\vspace{-.2in}
\bea
\!\!\begin{fmffile}{res2}
\Scale[0.5]{
\begin{gathered}
 \begin{fmfgraph*}(80,50)
    \fmfstraight
    \fmfleft{i1,i2}
    \fmfright{o1,o2}
    \fmf{photon}{i1,v1,i2} \fmf{photon}{o1,v2,o2}
    \fmf{gluon,left,tension=.5}{v1,v2,v1}
  \end{fmfgraph*}
  \end{gathered} 
} 
\end{fmffile} 
\!\!&\Rightarrow&\!\! -\fr12<\Big(\int V_{m2}\Big)^2>
= -\fr12<\Big[\int\fr14 (g^{\m\n}h^{\r\k}h_\k^{\s}  +g^{\r\s}h^{\m\k}h_\k^{\n}
 +h^{\m\n}h^{\r\s}\nn\\
 &&\hspace{1.5in}  -\fr14 g^{\m\n} g^{\r\s}h_{\k_1\k_2}h^{\k_1\k_2} ) ( \Ft_{\m\r}\Ft_{\n\s} )\Big]^2 >.
\eea
The cross-term diagram in Fig. 3 (d) is generated by the vacuum expectation value of the two vertices, one of which is the matter vertex $V_{m_2}$ and the other $V_{g}$. The diagram corresponds to
\bea
\hspace{-.1in}\begin{fmffile}{res3}
\Scale[0.5]{
\begin{gathered}
  \begin{fmfgraph*}(80,50)
    \fmfstraight
    \fmfleft{i1,i2}
    \fmfright{o}
    \fmf{photon}{i1,v1,i2} \fmf{gluon}{v2,o}
    \fmf{gluon,left,tension=.3}{v1,v2,v1}
  \end{fmfgraph*}
  \end{gathered} 
} 
\end{fmffile}  & \Rightarrow &   - <\!\int\! V_{m2}\!\int V_{g}\!>
= - <\!\int \Big(-\fr14\Big) \Big[g^{\m\n}h^{\r\k}h_\k^{\s}  +g^{\r\s}h^{\m\k}h_\k^{\n}
 +h^{\m\n}h^{\r\s}    \nn\\
 && \hspace{-.7in}-\fr14 g^{\m\n} g^{\m\n}h_{\k_1\k_2}h^{\k_1\k_2} 
\Big] \Big( \Ft_{\m\r}\Ft_{\n\s} \Big)  \times \fr1{\k'^2}\int \bigg\{ \Big(2g^{\b\b'}\tilde{\G}^{\a' \g\a}- g^{\a\b}\tilde{\G}^{\a' \g\b'}\Big)\pa_\g h_{\a\b}\, h_{\a'\b'} \nn\\
%%%%%%%%%%%%%%%%%%
&&\hspace{-.5in}\left.
+ \Big[\fr12(g^{\a\a'}g^{\b\b'}\vf^{\g\g'}+g^{\b\b'}g^{\g\g'}\vf^{\a\a'}
+g^{\a\a'}g^{\g\g'}\vf^{\b\b'})   -\fr12 g^{\g\g'}g^{\a'\b'}\vf^{\a\b}  \right.
\nn\\
&&\hspace{-.9in} \left.-\fr14 \vf^{\g\g'}g^{\a\b}g^{\a'\b'}
\Big] \pa_\g h_{\a\b}\, \pa_{\g'}h_{\a'\b'} +\Big( h_{\a\b}h_{\g\d}\Rt^{\a\g\b\d}-h_{\a\b}h^{\b}{}_\g \Rt^{\k\a\g}{}_{\k}
-\fr12 h^{\a\b}h_{\a\b}\Rt \Big) \right\}>.\nn\\
\eea
The computation of the diagrams with the heterotic loops serves as an example of the first-layer perturbation. For them it is necessary to use the full propagator in \rf{h2pt}, a step not needed for the other diagrams so far for a structural reason. In the first-layer perturbation, the graph to calculate is
\[
\begin{fmffile}{fullmixrel}
\Scale[0.99]{
 \begin{fmfgraph*}(90,70)\fmfpen{thick}
\fmfleft{i} \fmfright{o}
\fmf{photon,tension=4}{i,v1} \fmf{photon,tension=4}{v2,o}
\fmf{photon,left,tension=1}{v1,v2}  \fmf{gluon,left,tension=1}{v2,v1}
\end{fmfgraph*}
}
\end{fmffile}
\]
\vspace{-.4in}
\[
\mbox{Figure 5: first-layer perturbation diagram}
\]
Note that unlike Fig. 4 (a), the lines have been thickened. The external lines represent the full fields, i.e., the fields with tildes (see \rf{split}). By the same token, the internal lines represent the full propagators. (The two diagrams in Fig. 4 are the first two terms that result from, so to speak, $\vf_{\a\b}$-expanding the graph in Fig. 5. There are additional contributions coming from the internal lines when the full propagators are used.) As for the diagrams in Fig. 4, they can be set up in a manner similar to the others:
\bea
\hspace{-.2in}\begin{fmffile}{mixed2pt}
\Scale[0.6]{
\begin{gathered}
\begin{fmfgraph}(80,60)
\fmfleft{i} \fmfright{o}
\fmf{photon}{i,v1} \fmf{photon}{v2,o}
\fmf{photon,left,tension=.3}{v1,v2} \fmf{gluon,left,tension=.3}{v2,v1}
\end{fmfgraph}
  \end{gathered} 
} 
\end{fmffile} &\Rightarrow& -\fr12<\Big(\int V_{m3}\Big)^2>  = -\fr12<\Big(\int\fr12 (g^{\m\n}h^{\r\s}+g^{\r\s}h^{\m\n}) f_{\m\r}F_{\n\s}\Big)^2>
\nn\\
%%%%%%%%%%%%%%%%%%%
\begin{fmffile}{2ndmixrel}
\Scale[0.5]{
\begin{gathered}
 \begin{fmfgraph*}(80,60)
  \fmfleft{i1,i2,i3,i4}
    \fmfright{o}
 \fmf{phantom,tension=1}{i1,v1}   \fmf{photon,tension=1}{i2,v1}  \fmf{gluon,tension=1}{i3,v1} \fmf{phantom,tension=1}{i4,v1} 
\fmf{photon,tension=3}{v2,o}
\fmf{photon,left,tension=.5}{v1,v2}  \fmf{gluon,left,tension=.8}{v2,v1}
\end{fmfgraph*}
\end{gathered}
}
\end{fmffile}
&\Rightarrow& - <\int V_{m3}\int V_{m4}>= - <\int\fr12 (g^{\m\n}h^{\r\s}+g^{\r\s}h^{\m\n}) f_{\m\r}F_{\n\s}\nn\\
&&\hspace{1.0in}\times  \int \fr{-1}2 (\vf^{\m'\n'}h^{\r'\s'}+\vf^{\r'\s'}h^{\m'\n'}) f_{\m'\r'}F_{\n'\s'}>.
\eea
We will leave it for now and come back in section 3 where we show a more convenient way of effectively calculating all the contributions, including those arising from the full internal propagators.

%%%%%%%%%%%%%%%%%%%%%%%%%%%%%%%%%%%%%%%%%%%%
\subsection{vacuum-to-vacuum and tadpole diagrams}
%%%%%%%%%%%%%%%%%%%%%%%%%%%%%%%%%%%%%%%%%%%%

In the first-layer perturbation, the shifts in the cosmological and Newton's constants are caused by the vacuum-to-vacuum and tadpole diagrams, respectively (more details in section 4). 
 Fig. 6 below lists the diagrams for the pure gravity sector; there are similar diagrams for the matter-involving sector.  
For the graviton vacuum-to-vacuum amplitude, for example, one is to compute
\bea 
\int \prod_x dh_{\k_1\k_2}\;e^{\fr{i}{\k'^2} \int   \sqrt{-\gt}\,\Big( -\fr12\tilde{\N}_\g h^{\a\b}\tilde{\N}^\g h_{\a\b} \Big) }.
\eea
This vacuum energy amplitude in the first-layer perturbation will give a vacuum diagram and a tadpole diagram in the second-layer. 
As for the vector coupling renormalization, the diagram relevant is given in Fig. 7, a tadpole diagram with the graviton running on the loop.
The relevant first-layer vertex is
\bea
&& \hspace{-.1in}-\fr14\int \sqrt{-\gt}\Big[
\gt^{\m\n}h^{\r\k}h_\k^{\s}  +\gt^{\r\s}h^{\m\k}h_\k^{\n} -\fr12 \gt^{\m\n}hh^{\r\s}  -\fr12 \gt^{\r\s}hh^{\m\n}
+h^{\m\n}h^{\r\s}  \nn\\  
&&\hspace{1in}+\fr18 \gt^{\m\n} \gt^{\r\s}(h^2-2h_{\k_1\k_2}h^{\k_1\k_2} )
\Big] \Ft_{\m\r}\Ft_{\n\s}.   
\eea
The correlator to be computed is
\bea
&& 
 -\fr{i}{4\m^{2\e}e^2}\int \sqrt{-\gt} \Ft_{\m\r}\Ft_{\n\s}      \Big<
\gt^{\m\n}h^{\r\k}h_\k^{\s}  +\gt^{\r\s}h^{\m\k}h_\k^{\n} -\fr12 \gt^{\m\n}hh^{\r\s}  -\fr12 \gt^{\r\s}hh^{\m\n}
\nn\\  
&&\hspace{1.2in}+h^{\m\n}h^{\r\s} +\fr18 \gt^{\m\n} \gt^{\r\s}(h^2-2h_{\k_1\k_2}h^{\k_1\k_2} )
\Big>.  
\eea
With the self-contractions of the fluctuation fields, the correlator leads to a counterterm of the form $\sim \Ft_{\a\b}^2$.
Again the result vanishes due to the identity \rf{vmi} below in dimensional regularization. The shift can be introduced through finite renormalization. In section 4 below, we revisit the renormalization of the coupling constants by employing an alternate renormalization scheme where the cosmological constant is treated as a formal graviton mass.

\begin{figure}[t]
\begin{center}
\begin{fmffile}{vacandtad}
\parbox{30mm}{\begin{fmfgraph*}(75,50)
  %  \fmfstraight
%    \fmfleft{i1,i2}
   % \fmfright{o}
    %\fmf{photon}{i1,v1,i2} \fmf{gluon}{v2,o}
 %   \fmf{gluon,left,label=$h$,tension=.26}{v1,v2,v1}
   %   \fmflabel{A}{i1}  \fmflabel{A}{i2}  \fmflabel{$\vf$}{o}
   \fmfi{gluon}{reverse fullcircle scaled .5w shifted (.5w,.5h)}
  \end{fmfgraph*}\\

          \hspace{.38in}    (a)
 }  
 %%%%%%%%%%%%%%%%%%%%%%%%%%%%%%%%%%%%
\parbox{30mm}{\begin{fmfgraph*}(75,50)
  %  \fmfstraight
%    \fmfleft{i1,i2}
   % \fmfright{o}
    %\fmf{photon}{i1,v1,i2} \fmf{gluon}{v2,o}
 %   \fmf{gluon,left,label=$h$,tension=.26}{v1,v2,v1}
   %   \fmflabel{A}{i1}  \fmflabel{A}{i2}  \fmflabel{$\vf$}{o}
   \fmfi{dashes}{reverse fullcircle scaled .5w shifted (.5w,.5h)}
  \end{fmfgraph*}\\

          \hspace{.38in}    (b)
 }  
 %%%%%%%%%%%%%%%%%%%%%%%%%%%%%%%%%%%% 
 \parbox{30mm}{\begin{fmfgraph*}(75,50)
    \fmfleft{i}
    \fmfright{o}
    \fmf{gluon,tension=2}{i,v1}
    \fmf{gluon,left}{v1,o,v1}
  \end{fmfgraph*}\\

          \hspace{.6in}    (c)
 }  
%%%%%%%%%%%%%%%%%%%%%%%%%%%%%%%%%%%%
\quad\parbox{30mm}{\begin{fmfgraph*}(75,50)
    \fmfleft{i}
    \fmfright{o}
    \fmf{gluon,tension=2}{i,v1}
    \fmf{dashes,left}{v1,o,v1}
  \end{fmfgraph*}\\

          \hspace{.6in}    (d)
 }  
 
\end{fmffile}  
\vspace{.2in}
Figure 6: vacuum and tadpol diagrams
 \end{center} 
 %      \caption{\label{vtvt} vacuum and tadpol diagrams}
% \mbox{vacuum and tadpol diagrams}
\end{figure}

\begin{figure}[t]
\hspace{.3in}\begin{center}
    \Scale[.9]{
      \begin{fmffile}{gaugecoupling}
 %  \parbox{10mm}{
     \vspace{.8in}
     \hspace{1.3in}
    \begin{fmfgraph*}(70,100)\fmfpen{thick}
       \fmfleft{i}
       \fmfright{o}
       \fmftop{m}
       \fmfv{label=A,l.a=120}{i}
       \fmfv{label=A,l.a=60}{o}
       \fmflabel{$h$}{m}
       \fmf{photon,tension=.9}{i,v1}
       \fmf{photon,tension=.9}{v1,o}
       \fmf{gluon,left,tension=0.1}{v1,v1}
    \end{fmfgraph*}
%}
  \end{fmffile}
 \hspace{-2.5in}  \mbox{\large{{Figure 7.} vector coupling renormalization}}
    }
  \end{center} 
%\caption{ Figure 6: diagram relevant for vector coupling renormalization}
\end{figure}

%%%%%%%%%%%%%%%%%%%%%%%%%%%%%%%%%%%%%%%%
%%%%%%%%%%%%%%%%%%%%%%%%%%%%%%%%%%%%%%%%
\section{Flat space analysis}
%%%%%%%%%%%%%%%%%%%%%%%%%%%%%%%%%%%%%%%%
%%%%%%%%%%%%%%%%%%%%%%%%%%%%%%%%%%%%%%%%

 In this section we consider a flat background. The analysis can also be viewed as the computation of the divergences in a curved background:  the flat space analysis captures them since the ultraviolet divergence is a short-distance phenomenon. In the past the counter-terms were determined essentially by dimensional analysis and covariance \cite{Deser:1974cz}.\footnote{{Let us explain this point by taking the earlier work of \cite{'tHooft:1974bx} - which is simpler to see the point - whose methodology was shared by \cite{Deser:1974cz}. In the beginning of section 3 of \cite{'tHooft:1974bx}, the authors considered a scalar system in a flat background. The action (3.1) therein contains external fields; the 1PI effective action was worked out in the usual way. Note that the background fields, namely the external fields, are completely offshell. Afterwards they considered a gravity case whose action is given in (3.8). This time, however, the form of the action (3.9) was fixed based on the covariance and dimensional analysis. The coefficients of the counterterms were then determined by considering certain Feynman diagrams with {\em onshell} background fields {and using the usual traceful propagator}. {As heavily stressed in our previous works \cite{Park:2014noa,Park:2015ota}, only the traceless propagator maintains the covariance.} In the present work, the analogous calculation is performed with offshell background fields, just as in the first example, i.e., the scalar case of \cite{'tHooft:1974bx}.} \la{fnt9}} We directly calculate them in the refined background field method; dimensional analysis and covariance play the {\em subsidiary} role of checking the results.

Let us consider a flat background so that
\bea
\gh_{\m\n}\equiv  h_{\m\n}+\tilde{g}_{\m\n}\quad,\quad
\Ah_\m \equiv a_\m+\At_\m
  \la{gshiftq}
\eea
where
\bea
\quad \tilde{g}_{\m\n} \equiv \vf_{\m\n}+g_{\m\n}\quad,\quad \At_\m \equiv A_\m+A_{0\m} \la{splitq}
\eea
and now
\bea
g_{\m\n}=\eta_{\m\n}\quad ,\quad A_{0\m}=0.
\eea
We employ dimensional regularization. 
In what follows we will present the explicit flat spacetime computations for the two-point diagrams considered for a generic background $g_{\m\n}$ in the previous section. Although the techniques of the counter-term computation themselves are similar to those used in the pure gravity \cite{Park:2015ota} and gravity-scalar \cite{Park:2016zgt} analyses, the present case has several additional complications. As an unexpected spin-off of our direct approach, we will see how the newly noted gauge choice-dependence issue is resolved in the present framework. {In section 3.2 we also address how the well-known gauge-choice dependence may be avoided - especially for a de Donder type gauge - for the physical states in the present framework.}

%%%%%%%%%%%%%%%%%%%%%%%%%%%%%%%%%%%%%%%%
\subsection{two-point diagrams}
%%%%%%%%%%%%%%%%%%%%%%%%%%%%%%%%%%%%%%%%

The pure gravity sector was analyzed in \cite{Park:2015ota}. Consider the ghost loop diagram in Fig. \ref{fig:2gh} (b) first. The ghost vertex takes, in the flat spacetime,
\[
V_{C}= -\Big[
-\tilde{\G}^\l_{\m\n}({ -C_\l} \pa^\m \bar{C}^\n+\bar{C}_\l\pa^\m {C}^\n  )
 -(\eta^{\n\b}\vf^{\m\a}+\eta^{\m\a}\vf^{\n\b})\pa_\b \bar{C}_\a \pa_\n {C}_\m
\Big] {+}R_{\m\n}\bar{C}^\m C^\n.  \nn\\  \la{ghkinexp}
\]
Let us define, for convenience,
\bea
V_{C}=V_{C,I}+V_{C,II}
\eea
with
\bea
&&\hspace{-.5in} V_{C,I} \equiv -\Big[
-\tilde{\G}^\l_{\m\n}({ -C_\l} \pa^\m \bar{C}^\n+\bar{C}_\l\pa^\m {C}^\n  ) -(\eta^{\n\b}\vf^{\m\a}+\eta^{\m\a}\vf^{\n\b})\pa_\b \bar{C}_\a \pa_\n {C}_\m  \Big] \nn\\
&&\hspace{1.6in}  V_{C,II} \equiv R_{\m\n}\bar{C}^\m C^\n.
\eea
The correlator to be computed is 
\bea
 &&\hspace{-.3in} -\fr12 { \fr{1}{\k'^4}}<\Big(\int V_{C,I}+V_{C,II} \Big)^2> = -\fr12 { \fr{1}{\k'^4}}<\Big\{\int \Big[
-\tilde{\G}^\l_{\m\n}(\pa^\m \bar{C}^\n { C_\l} -\pa^\m {C}^\n  \bar{C}_\l) \nn\\
&&\hspace{.7in} -(\eta^{\n\b}\vf^{\m\a}+\eta^{\m\a}\vf^{\n\b})\pa_\b \bar{C}_\a \pa_\n {C}_\m
\Big]  -R_{\m\n}\bar{C}^\m C^\n\Big\}^2>.
  \la{totgh1}
\eea
To see how the dimensional analysis and covariance can be utilized to check the final results, consider, e.g., $<(\int V_{C,I})^2>$; a direct calculation yields
\bea
\hspace{-.2in} -\fr12 { \fr{1}{\k'^4}}<\Big(\int V_{C,I}\Big)^2> 
\eea
\vspace{-.1in}
\[
= -\fr12 \fr{\G(\ve)}{(4\pi)^2}\int \Big[ -\fr{2}{15}\pa^2\vf_{\m\n}\pa^2 \vf^{\m\n}+\fr{4}{15}\pa^2 \vf^{\a\k}\pa_\k \pa_\s \vf_\a^\s
-\fr{1}{30}(\pa_\a \pa_\b \vf^{\a\b})^2
\Big] 
\]
where the parameter $\ve$ is related to the total spacetime dimension $D$ by
\bea
D=4-2\ve.
\eea	
The result above ({and some below}) were obtained with the help of the Mathematica package xAct`xTensor` in performing the index contractions. By invoking dimensional analysis and covariance, one expects the result to come out to be a sum of $R^2$ and $R_{\m\n}^2$ to the second order of $\vf_{\r\s}$ with appropriate coefficients. With the traceless condition $\vf=0$ explicitly enforced, $R^2$ and $R_{\m\n}^2$ are given, to the second order in $\vf_{\a\b}$, by
\bea
R^2 &=& \pa_{\m}\pa_{\n}\vf^{\m\n}\,\pa_{\r}\pa_{\s}\vf^{\r\s}
\nn\\
R_{\a\b}R^{\a\b} &=& \fr14\Big[\pa^2 \vf^{\m\n}\,\pa^2 \vf_{\m\n}-2\pa^2 \vf^{\a\k}\pa_\k \pa_\s \vf_\a^\s
+2(\pa_{\m}\pa_{\n}\vf^{\m\n})^2
 \Big];
\la{covctr}
\eea
from these it follows 
\bea
-\fr12 { \fr{1}{\k'^4}}<\Big(\int V_{C,I}\Big)^2> = -\fr1{2} \fr{\G(\ve)}{(4\pi)^2}\int \Big[-\fr{8}{15}\Rt_{\a\b}\Rt^{\a\b}+\fr{7}{30}\Rt^2\Big].
\eea
The tildes on the fields in the counter-terms will be omitted from now on. Let us complete the other terms in \rf{totgh1}; collecting all, one gets, for the ghost diagram,
\bea
\begin{fmffile}{pure2}
\!\!\Scale[0.4]{
\begin{gathered}
 \begin{fmfgraph*}(80,50)
\fmfleft{i} \fmfright{o}
\fmf{gluon,tension=6}{i,v1} \fmf{gluon,tension=6}{v2,o}
\fmf{dashes,left,tension=1}{v1,v2,v1}
\end{fmfgraph*}
\end{gathered}
}
\end{fmffile}  
%%%%%%%%%%%%%%%%%%%%%%%%%%%%%%%%%%%%%%%%%%%%%%%%%
%%%%%%%%%%%%%%%%%%%%%%%%%%%%%%%%%%%%%%%%%%%%%%%%
&\Rightarrow& -\fr12 \fr{\G(\ve)}{(4\pi)^2}\int \Big[ \fr{7}{15}R_{\m\n}{ R^{\m\n}} {+\fr{17}{30}}R^2
\Big]. \la{tgh}
\eea
As for the graviton-loop diagram in Fig. \ref{fig:2gh} (a), the vertex $V_{g} $ takes
\bea
 &&\hspace{-.41in} V_{g} \equiv  \Big(2\h^{\b\b'}\tilde{\G}^{\a' \g\a}\!-\! \h^{\a\b}\tilde{\G}^{\a' \g\b'}\Big)\pa_\g h_{\a\b}\, h_{\a'\b'}  
 \!+\! \Big[\fr12(\h^{\a\a'}\h^{\b\b'}\vf^{\g\g'}+\! \h^{\b\b'}\h^{\g\g'}\vf^{\a\a'}   \nn\\
&&\hspace{.5in} +\h^{\a\a'}\h^{\g\g'}\vf^{\b\b'})  -\fr12 \h^{\g\g'}\h^{\a'\b'}\vf^{\a\b}  
-\fr14 \vf^{\g\g'}\h^{\a\b}\h^{\a'\b'}
\Big] \pa_\g h_{\a\b}\, \pa_{\g'}h_{\a'\b'} \nn\\
&& \hspace{.5in}+ \Big( h_{\a\b}h_{\g\d}\Rt^{\a\g\b\d}-h_{\a\b}h^{\b}{}_\g \Rt^{\k\a\g}{}_{\k}
-\fr12 h^{\a\b}h_{\a\b}\Rt
 \Big).
\eea
{Let us define:
\bea
V_{g,I} &\equiv&   \Big(2\eta^{\b\b'}\tilde{\G}^{\a' \g\a}- \eta^{\a\b}\tilde{\G}^{\a' \g\b'}\Big)\pa_\g h_{\a\b}\, h_{\a'\b'}  \nn\\
V_{g,II} &\equiv& \Big[\fr12(\eta^{\a\a'}\eta^{\b\b'}\vf^{\g\g'}+\eta^{\b\b'}\eta^{\g\g'}\vf^{\a\a'}
+\eta^{\a\a'}\eta^{\g\g'}\vf^{\b\b'})\nn\\
&&-\fr14 \vf\, \eta^{\a\a'}\eta^{\b\b'}\eta^{\g\g'}-\fr12 \eta^{\g\g'}\eta^{\a'\b'}\vf^{\a\b}  \nn\\
&&+\fr14 (-\vf^{\g\g'}+\fr12 \vf \eta^{\g\g'})\eta^{\a\b}\eta^{\a'\b'}
\Big] \pa_\g h_{\a\b}\, \pa_{\g'}h_{\a'\b'}  \la{lv12q}
\eea
\bea
\hspace{-.2in}{V_{g,III}} = \sqrt{-\gt}\Big( h_{\a\b}h_{\g\d}\Rt^{\a\g\b\d}-h_{\a\b}h^{\b}{}_\g \Rt^{\k\a\g}{}_{\k}
-\fr12 h^{\a\b}h_{\a\b}\Rt
\Big).  \la{gverq}  
\eea
}
By using the traceless propagator one can show:
\bea
\begin{fmffile}{pure1}
\Scale[0.4]{
\begin{gathered}
 \begin{fmfgraph*}(80,50)
\fmfleft{i} \fmfright{o}
\fmf{gluon,tension=6}{i,v1} \fmf{gluon,tension=6}{v2,o}
\fmf{gluon,left,tension=1}{v1,v2,v1}
\end{fmfgraph*}
\end{gathered}
}
\end{fmffile}
&\Rightarrow &  -\fr12\fr{\G(\ve)}{(4\pi)^2}\int \Big[-\fr{23}{20}R_{\m\n}R^{\m\n}-{ \fr{23}{40}}R^2\Big]. 
\la{tghpri}
\eea
The correlators for the matter-involving sector have also been outlined in the previous section. Their flat spacetime evaluation leads to the following results for the diagrams in Fig. 3 (a)-(c): 
\bea
 \begin{fmffile}{res1}
\Scale[0.5]{
\begin{gathered}
\begin{fmfgraph}(80,50)
\fmfleft{i} \fmfright{o}
\fmf{gluon}{i,v1} \fmf{gluon}{v2,o}
\fmf{photon,left,tension=.3}{v1,v2,v1}
\end{fmfgraph}
  \end{gathered} 
} 
\end{fmffile} &\Rightarrow& 
\fr{\G(\ve)}{(4\pi)^2}\int \Big(\fr1{30} R^2-\fr1{10}R_{\a\b}R^{\a\b} \Big)  \nn\\
%%%%%%%%%%%%%%%%%%%%%%%%%%%%%%%%%%%%%%
\begin{fmffile}{respl}
\Scale[0.5]{
\begin{gathered}
\begin{fmfgraph}(80,50)
\fmfleft{i} \fmfright{o}
\fmf{gluon}{i,v1} \fmf{gluon}{v2,o}
\fmf{dots,left,tension=.3}{v1,v2,v1}
\end{fmfgraph}
  \end{gathered} 
} 
\end{fmffile} &\Rightarrow& 
-\fr{\G(\ve)}{(4\pi)^2} \fr1{15}\int R_{\a\b}R^{\a\b}   \nn\\
%%%%%%%%%%%%%%%%%%%%%%%%%%%%%%%%%%%%%%
 \begin{fmffile}{res2}
\Scale[0.5]{
\begin{gathered}
 \begin{fmfgraph*}(80,50)
    \fmfstraight
    \fmfleft{i1,i2}
    \fmfright{o1,o2}
    \fmf{photon}{i1,v1,i2} \fmf{photon}{o1,v2,o2}
    \fmf{gluon,left,tension=.5}{v1,v2,v1}
  \end{fmfgraph*}
  \end{gathered} 
} 
\end{fmffile} 
&\Rightarrow &  \fr{{ \k'^4}\,\G(\ve)}{(4\pi)^2} \fr3{64} \int (F_{\a\b}F^{\a\b})^2.
\eea
These results are covariant as expected.
The direct calculation of the diagram in Fig. 3 (d) yields
\bea
\begin{fmffile}{res3}
\Scale[0.5]{
\begin{gathered}
  \begin{fmfgraph*}(80,50)
    \fmfstraight
    \fmfleft{i1,i2}
    \fmfright{o}
    \fmf{photon}{i1,v1,i2} \fmf{gluon}{v2,o}
    \fmf{gluon,left,tension=.3}{v1,v2,v1}
  \end{fmfgraph*}
  \end{gathered} 
} 
\end{fmffile}\bigg{|}_{{V_{g,I}+V_{g,II}}}   \Rightarrow    
 \fr{\k'^2\,\G(\ve)}{(4\pi)^2}  \int \Big(\fr1{16}F_{\m\n}F^{\m\n} \pa_\a\pa_\b \vf^{\a\b} + \fr12 F_{\m\k}F_\n{}^{\k} \pa^2 \vf^{\m\n} \Big)
 \la{f2c}
\eea
which is non-covariant.\footnote{In the case of the Einstein-scalar system analyzed in \cite{Park:2016zgt}, we obtained a covariant result for a similar diagram,
\begin{fmffile}{gl2s}
\Scale[0.3]{
\begin{gathered}
  \begin{fmfgraph*}(80,50)
    \fmfstraight
    \fmfleft{i1,i2}
    \fmfright{o}
    \fmf{plain}{i1,v1,i2} \fmf{gluon}{v2,o}
    \fmf{gluon,left,tension=.3}{v1,v2,v1}
  \end{fmfgraph*}
  \end{gathered} 
\la{ftn10}} 
\end{fmffile}. The present non-covariant result is one reflection of the complexity of the gauge matter system.} As a matter of fact, this is the diagram that suggests the solution for the gauge choice-dependence. This non-covariant result will be examined in section 3.2 and we will see how the covariance is restored. 
The diagram above also receives a contribution from ${V_{g,III}}$ vertex:
\bea
 \begin{fmffile}{res3}
\Scale[0.5]{
\begin{gathered}
  \begin{fmfgraph*}(80,50)
    \fmfstraight
    \fmfleft{i1,i2}
    \fmfright{o}
    \fmf{photon}{i1,v1,i2} \fmf{gluon}{v2,o}
    \fmf{gluon,left,tension=.3}{v1,v2,v1}
  \end{fmfgraph*}
  \end{gathered} 
} 
\end{fmffile}\bigg{|}_{{V_{g,III}}} &\Rightarrow&    
  \fr{\k'^2\,\G(\ve)}{(4\pi)^2} \int \Big(\fr34 F_{\a\k}F_\b{}^{\k}R^{\a\b}+\fr18 F_{\a\b}F^{\a\b}R \nn\\
&&  +\fr14 F_{\a\d}F_{\b\g}R^{\a \b\g\d} -\fr14 F_{\a\b}F_{\g\d}R^{\a \b\g\d}\Big). 
\eea
As for the diagrams with the inhomogeneous loops, the first-layer diagram to be computed is the one in Fig. 5. It corresponds to several second-layer diagrams, two of which are Fig. 4 (a) and (b); one can show
\bea
\hspace{-.2in}\begin{fmffile}{mixed2pt}
\Scale[0.6]{
\begin{gathered}
\begin{fmfgraph}(80,60)
\fmfleft{i} \fmfright{o}
\fmf{photon}{i,v1} \fmf{photon}{v2,o}
\fmf{photon,left,tension=.3}{v1,v2} \fmf{gluon,left,tension=.3}{v2,v1}
\end{fmfgraph}
  \end{gathered} 
} 
\end{fmffile} &\Rightarrow&  \fr{{ \k'^2}}2   \fr{\G(\ve)}{(4\pi)^2}\int \Big(\fr13 \pa_\a F^\a{}_{\k} \pa_\b F^{\b\k}-\fr1{12} \pa_\r F_{\a\b}\pa^\r F^{\a\b} \Big)   \nn\\
%%%%%%%%%%%%%%%%%%
\begin{fmffile}{2ndmixrel}
\Scale[0.5]{
\begin{gathered}
 \begin{fmfgraph*}(80,60)
  \fmfleft{i1,i2,i3,i4}
    \fmfright{o}
 \fmf{phantom,tension=1}{i1,v1}   \fmf{photon,tension=1}{i2,v1}  \fmf{gluon,tension=1}{i3,v1} \fmf{phantom,tension=1}{i4,v1} 
\fmf{photon,tension=3}{v2,o}
\fmf{photon,left,tension=.5}{v1,v2}  \fmf{gluon,left,tension=.8}{v2,v1}
\end{fmfgraph*}
\end{gathered}
}
\end{fmffile}
&\Rightarrow & {\k'^2}\fr{\G(\ve)}{(4\pi)^2} \int \Big( \fr13 F_{\a\k}\pa_\l \pa^\b F_\b{}^\k \vf^{\a\l}- \fr1{12} F_{\a\k}\pa^2F_\b{}^\k \vf^{\a\b} \Big)
\la{ild3}
%%%%%%%%%%%%%%%%%%%%%%%%%%%
\eea 
where all of the index contractions are done with the flat metric. Whereas the first diagram is covariant at the leading order, the second diagram is not at its given order, the $\vf_{\a\b}$-linear order. There are also contributions arising from the higher-order internal propagators and all of these three different contributions are required for the covariance since they altogether correspond to the single first-layer diagram in Fig. 5. Keeping track of the higher-order internal propagators obviously requires the full (or at least higher-order) propagator expression $\tilde{\D}$.  Therefore, instead of separately computing the individual contributions, it will be more economical to compute them at one stroke. The calculation can be done by performing the following steps: let us consider   
\bea
{\bf V}\equiv \fr12  \Big[ \gt^{\r\s}h^{\m\n}+\gt^{\m\n}h^{\r\s}\Big] f_{\m\r}F_{\n\s}
\eea
where the contractions are carried out by $\gt_{\m\n}$. 
At this point we introduce the orthonormal basis $e_a^\m$:
\be
\et_a^\m \et_b^\n \gt_{\m\n}=\h_{ab} \la{bitf}
\ee
where the Latin indices run $a,b=0,1,2,3$. The full scalar propagator $\tilde{\D}$ can be written 
\be
\tilde{\D}(X_1-X_2)=\int \fr{d^4L}{(2\pi)^4}\fr{e^{iL_c (X_1-X_2)^c}}{i L_a L_b \h^{ab}}  \label{thickprop}
\ee
where $X^a$ and $L_c$ are the coordinates and momenta associated with the orthonormal basis. 
Then the computation of the two-point amplitude goes identically with that of Fig. 4 (a); switching back to the original frame, one gets
\bea
\hspace{-.2in}\begin{fmffile}{fpdwl}
\Scale[0.6]{
\begin{gathered}
\begin{fmfgraph}(80,60)\fmfpen{thick}
\fmfleft{i} \fmfright{o}
\fmf{photon}{i,v1} \fmf{photon}{v2,o}
\fmf{photon,left,tension=.3}{v1,v2} \fmf{gluon,left,tension=.3}{v2,v1}
\fmflabel{A}{i}   \fmflabel{$\gt$}{o}
\end{fmfgraph}
  \end{gathered} 
} 
\end{fmffile} \Rightarrow  -\fr12<\Big(\int {\bf V}\Big)^2>=\fr{{ \k'^2}}2   \fr{\G(\ve)}{(4\pi)^2}\int \Big( \fr13 \nabla_\a F^\a{}_{\k} \nabla_\b F^{\b\k}-\fr1{12} \nabla_\r F_{\a\b}\nabla^\r F^{\a\b} \Big). \la{thickdia} \nn\\
%%%%%%%%%%%%%%%%%%
\eea
The analysis of the vacuum-to-vacuum amplitudes and tadpoles will be presented in section 4.

%%%%%%%%%%%%%%%%%%%%%%%%%%%%%%%%%%%%%%%%
\subsection{on gauge-choice- and background- independence \la{sogbi}}
%%%%%%%%%%%%%%%%%%%%%%%%%%%%%%%%%%%%%%%%

Above, we have evaluated the counter-terms for the diagrams in {Fig.} 2 to 6, and they have led to different types of the counter-terms, one of which, i.e., eq. \rf{f2c}, has come out non-covariant. This means that the effective action, as it stands, is non-covariant and gauge fixing-dependent.\footnote{In an intensive work of \cite{Odintsov:1991fk} (see also \cite{Odintsov:1989gz}), a certain gauge-choice dependence was found even though the Vilkovisky's method \cite{Vilkovisky:1984st} was employed. (A similar related observation was made in \cite{Falls:2015qga}.) {The gauge-choice dependence newly found in the present framework should have the same origin as that of \cite{Odintsov:1991fk} (see below) but be a different manifestation. The gauge-choice dependence in the present work occurs through breaking of the covariance. It is milder and is easily fixable as we will see.} The well-known gauge-choice dependence {- which is addressed at the end of this subsection -} is that the coefficients of the covariant terms in the 1PI effective action depend, in general, on the gauge of one's choice (see, e.g., \cite{Modesto:2017hzl} and \cite{Falls:2015qga}). Whereas this type of gauge-dependence occurs to a YM theory as well, the present gauge-dependence is unique to a gravity theory. \la{ftn13}
} It turns out that these two problems have the following common solution: once the gauge-fixing\footnote{Note that the gauge-fixing 
\bea
\pa_\m \vf^{\m\n}-\pa^\n \vf =0  \la{vfgf}
\eea
reduces to \rf{vfgfr} once the traceless condition $\vf=0$ is enforced.} 
\bea
\pa_\m \vf^{\m\n}=0 \la{vfgfr}
\eea
is explicitly imposed on the effective action, the covariance and gauge-choice independence are restored.

To see this, let us examine the non-covariant counter-terms for Fig. 3 (c) given in \rf{f2c}. Note that the first term in \rf{f2c} vanishes upon imposing the strong form of the gauge condition $\pa_\m \vf^{\m\n}=0$,
which implies\footnote{It is with the following caveat. Since the scalar curvature $R$ is given  by
 \bea
 R=\pa_\a\pa_\b \vf^{\a\b}
 \eea
 to the linear order, it is not possible, with the strong form of the gauge condition, to probe the presence of the $R$-factor through the current linear-order calculation. For that, it is necessary to go to the second order. 
 }
 \bea
 \pa_\n\pa_\m \vf^{\m\n}=0;
 \eea
with it eq. \rf{f2c} now takes
 \bea
\;\;\begin{fmffile}{res3}
\Scale[0.5]{
\begin{gathered}
  \begin{fmfgraph*}(80,50)
    \fmfstraight
    \fmfleft{i1,i2}
    \fmfright{o}
    \fmf{photon}{i1,v1,i2} \fmf{gluon}{v2,o}
    \fmf{gluon,left,tension=.3}{v1,v2,v1}
  \end{fmfgraph*}
  \end{gathered} 
} 
\end{fmffile}\bigg{|}_{{V_{g,I}+V_{g,II}}}   \Rightarrow    
 \fr{\k'^2\,\G(\ve)}{(4\pi)^2}  \int \Big( \fr12 F_{\m\k}F_\n{}^{\k} \pa^2 \vf^{\m\n} \Big)= -\fr{\k'^2\,\G(\ve)}{(4\pi)^2}  \int  F_{\m\k}F_\n{}^{\k}  R^{\m\n} \nn\\  \la{f2c2}
\eea
where the second equality is valid, as usual, up to a certain order of $\vf_{\a\b}$, the linear order for the present case. 
Note that above, the following identity at $\vf_{\r\s}$-linear order has been used:
\bea
R_{\m\n}&=& \fr12 (\pa^\k\pa_\m \vf_{\k\n}+\pa^\k\pa_\n \vf_{\k\m}-\pa_\m\pa_\n \vf-\pa^2 \vf_{\m\n} ) 
             = - \fr12  \pa^2 \vf_{\m\n} 
\eea
where the second equality results once the gauge conditions are enforced.

%One is now in a position to address the longstanding issue of the well-known gauge choice-dependence of the effective action. In the present framework, 
The effective action becomes fully covariant and gauge-choice independent after enforcing $\pa_\n \vf^{\m\n}=0$. In this sense, the novel gauge-choice dependence found in the present work is mild, due to the use of the traceless propagator and refined background field method. Also, among the terms explicitly evaluated above, only \rf{f2c} has the issue; all the other terms are gauge-fixing independent. In particular, the $F^2$ is gauge-choice independent although we did not record the result (since our focus is the cosmological and Newton's constants), a result consistent with \cite{Huggins:1987zw}.

One should view the covariant action as still supplemented by the gauge-fixing. (This is just like the classical action:  the classical action is fully covariant but is to be supplemented by a gauge-fixing condition.) If one chooses a different gauge-fixing and carries out the amplitude computations in that gauge, one should get exactly the same covariant effective action up to the terms that can be removed by that gauge-condition; this time, the action is supplemented with the very gauge-fixing condition that one has chosen. Therefore, the gauge-choice independence of the effective action should be interpreted to mean that the action is covariant after enforcing the strong form of the gauge condition and that the covariant action is to be supplemented by the gauge-fixing condition of one's choice. (But one can of course choose any gauge-fixing (namely even a gauge-fixing different from the initial one) once the covariant effective action is obtained.)

One may wonder whether the appearance of the factors of $\pa_\m \vf^{\m\n}$ in the counter-term calculation of \rf{f2c} could by any chance be made to disappear, say, without imposing the gauge condition. It appears that the gauge choice-dependence has a deeper root. To be specific, let us consider the proof of the gauge-choice independence in chap. 15 of \cite{Weinberg2}. The proof is for a gauge theory in the ordinary (i.e., non-BFM) path-integral. The gauge choice-dependence gets to reside in a field-independent constant (therein denoted by $C$; see eq. (15.5.19)), which is then duly disregarded. If one employs the BFM, however, that constant comes to depend on the background fields (say, $\vf_{\m\n}$ for the present case, for example) and this must be the gauge choice-dependence that we have observed. This shows that the BFM, refined or not, has a limitation when applied to a gravitational system: it is introduced aiming for a more covariant treatment of the effective action computation. It turns out to be at odds with the gauge-choice independence.\footnote{Nevertheless, the refined BFM has an advantage compared to the conventional BFM in that the latter would yield results non-covariant in an uncontrollable way whereas the former gives the results covariant up to the gauge-choice dependent terms that can be removed by enforcing the strong form of the gauge condition.} The limitation is overcome by imposing the gauge condition in its strong form as we have just discussed.

{Finally, let us address the well-known longstanding gauge-choice dependence followed by background (in)dependence. This long-noted gauge choice-dependence of the effective action had been studied in a number of papers including \cite{Vilkovisky:1984st,Fradkin:1983nw,Modesto:2017hzl,Huggins:1987zw,Toms:2008dq,Odintsov:1989gz,Odintsov:1991fk,Falls:2015qga} in the past. In \cite{Falls:2015qga}, in particular, it was shown to disappear onshell. We examine two aspects of the issue. The first is most relevant for effective action computation and beta function computation in section 4.3 below. The second concerns the gauge-choice independence of scattering amplitudes, i.e., the gauge-choice independence of S-matrix.
	
For the first aspect, let us recall that one specific way of posing the potential gauge-choice dependence issue is to consider the following type of a gauge-fixing term,
\be
-\fr1{2\xi}\Big[\tilde{\nabla}_\n h^{\m\n}-\fr{\r}2 \tilde{\nabla}^\m h \Big]^2  
\ee	
and ask whether or not the constants $\xi$, $\r$ appear in the effective action. Since the propagator is traceless, the trace term $h$ can simply be omitted, as commented earlier. This ensures $\r$-independence of the computation. As for the constant $\xi$, see the comments below \rf{scamp}.
%the traceless condition is only satisfied for $\a=1$. In other words, other values of $\a$ would lead to a propagator that contains a trace piece, and will thus lead to non-covariant results.   

The second aspect will be most relevant for the gauge-choice independence of the S-matrix. The upshot is that the dependence may be avoided for the ``physical states." The present framework is in line with \cite{Falls:2015qga}, and refines it in the sense that only a part of the field equations should be required. To see this, let us recall the the physical states are defined as a solution of the Hamiltonian and momentum constraints in the ADM formalism; see, e.g., \cite{Park:2019amz} for more details. The full nonlinear form of the de Donder gauge is given by
\be
\gh^{\r\s}\hat{\G}^\m_{\r\s}=0 \la{dDgauge}
\ee
where $\gh^{\r\s}$ denotes a generic metric and $\hat{\G}^\m_{\r\s}$ the Christoffel symbols. In the ADM formalism the condition eq. \rf{dDgauge} splits into
\bea
&& \hspace{.5in}(\pa_{x^3}-\Nh^m \pa_m) \nh=\nh^2\Kh  \nn\\
&& (\pa_{x^3}-\Nh^n \pa_n)\Nh^m=\nh^2(\hat{\g}^{mn}\pa_n \ln \nh-\hat{\g}^{pq}\hat{\G}^m_{pq})
\la{ADMddq}
\eea
where $\nh,\Nh^m$ are the lapse function and shift vector; $\Kh$ denotes trace of the second fundamental form; $\hat{\g}^{pq}$ is the 3D metric. Let us demonstrate the point for the fluctuations in a flat background; the conclusion will remain valid for a curved background. The physical states are defined as a solution of the Hamiltonian and momentum constraints. Since $\nh,\Nh^m$ are non-dynamical, they can be gauged away. (See \cite{Park:2019amz} for a review.) With this the first equation of \rf{ADMddq} is satisfied, and the second becomes
\bea
\hat{\g}^{pq}\hat{\G}^m_{pq}=0 \la{3DdD}
\eea
which is nothing but the gauge-fixing condition adopted for the 3D analysis \cite{Park:2014noa}.
This suggests, in light of the discussion above of the origin of the dependence, that the amplitude will be independent of the non-physical states, i.e., will be gauge-fixing-independent, once the external states are restricted to the physical states. Recall now that the Hamiltonian and momentum constraints are nothing but the field equations of the lapse and shift in the Lagrangian formalism \cite{Park:2014noa}. In other words, they are partially onshell (but don't have to be fully onshell). More specifically, let us consider a scattering amplitude
\bea
<out;\{\b\}|in;\{\a\}> \la{scamp}
\eea
where $|in;\{\a\}>$ and $|out;\{\b\}>$ denote in- and out- states, respectively; $\{\a\},\{\b\}$ collectively stand for various quantum numbers - such as angular momentum, spin, etc - carried by the Fock oscillators. Since the metric field is not a gauge singlet, such an amplitude, not being gauge invariant, will be generally gauge choice-dependent. By definition the physical states are inert under the bulk gauge transformation: once the external states are taken to be the physical states,\footnote{At the action level, taking the physical vacuum states corresponds to carrying out certain dimensional reduction whose explicit procedure can be found in \cite{Park:2018xtt} for pure gravity. The reduction process is of course what makes the action partially onshell \cite{Park:2019amz} as mentioned above \rf{scamp}.} the amplitude will be gauge invariant and thus independent of gauge redundancy. Similarly, as for the potential $\xi$-dependence of the effective action, it is expected that the effective action will, in general, depend on $\xi$. This is because the effective action corresponds to the amplitude \rf{scamp}, with only the 1PI diagrams collected, with both in- and out- states taken as the 4D vacuum, and the 4D vacuum is not the physical vacuum. Once the in- and out states are taken as the physical vacuum, the resulting effective action should be independent of gauge redundancy. However, a subtle possibility is that the 3D de Donder gauge \rf{3DdD} may also contain, when considered in the context of a large gauge transformation, labeling different sectors of the physical degrees of freedom. (The reason for this expectation is
as follows. In \cite{Park:2018xtt} it was shown that the residual symmetry of the de Donder
gauge imposes a constraint on the gauge parameter. It is unlikely that the
parameters of the large gauge transformation will satisfy it.) If that's the case, which seems to be a reasonable possibility, even the reduced action (see footnote 13) will depend on $\xi$ but the interpretation would be different: the dependence should be viewed as analogous to dependence of an amplitude on a global symmetry.

As for the background independence, the use of the full propagator greatly reduces the amount of computation, as we have seen in \rf{thickdia}. Combined with freedom in choosing renormalization conditions, the use of the first-layer or `one-stroke' propagator may also have far-reaching consequences in implementing background-independent analysis. It is evident that computations carried out with the full propagator \rf{thickprop} will yield background-independent results. The background dependence is implicit in the transformation associated with \rf{bitf}. As for the infrared sector contribution, at the end it may well be that the ever-powerful freedom in the renormalization conditions can be chosen to reflect the characteristics of the actual background.

%%%%%%%%%%%%%%%%%%%%%%%%%%%%%%%%%%%%%%%%%%%%
%%%%%%%%%%%%%%%%%%%%%%%%%%%%%%%%%%%%%%%%%%%%
\section{One-loop renormalization}
%%%%%%%%%%%%%%%%%%%%%%%%%%%%%%%%%%%%%%%%%%%%
%%%%%%%%%%%%%%%%%%%%%%%%%%%%%%%%%%%%%%%%%%%%

The focus of the previous section was on the divergent parts of the diagrams; the analysis was involved but relatively straightforward. The forms of the counter-terms have been obtained with the infinite parts of the coefficients specified. As well known in the standard quantum field theory, one has the freedom to adjust the finite parts of the coefficients through the renormalization scheme, which one may take to be the modified minimal subtraction $\overline{\mbox{MS}}$ (see, e.g., \cite{Sterman} for a review). The main focus of the present section is to carry out the renormalization in detail and study its implications.

For the detailed analysis involving the renormalization conditions, it is convenient, as commonly done, to introduce a scale parameter $\m$ by making the following scaling:
\bea
\k^2\ra \m^{{-D+4}}\k^2.
\eea
With this, eq.\!\! \rf{EM} takes
\bea
S=\int \sqrt{-\gh}\;\Big(\fr1{\k^2 \m^{{4- D}}}\Rh-\fr14 \Fh_{\m\n}^2 \Big).
\eea
One can proceed and compute various amplitudes and counter-terms; that was basically what we did in the previous section, but this time the parameter $\m$ will be included. The finite parts can now be kept track of with the fixed renormalization scheme. 

One of the main goals of this section is to analyze the renormalization of the cosmological and Newton's constants (earlier works can be found, e.g., in \cite{Reuter:1996cp},\cite{Donkin:2012ud} and \cite{Falls:2015qga}.) In its entirety the procedure involves dealing with an infinite number of counter-terms, not just the cosmological constant and Einstein-Hilbert terms. 
 It will be nevertheless useful to first hone in on the renormalization of those two constants, a task undertaken at the end of section 4.1 before working out the further details of the whole procedure in section 4.2. This is because these constants carry special physical meanings unlike the other newly appearing couplings that will ultimately be absorbed by a metric field redefinition. Moreover, there are some subtleties in the evaluation of the diagrams responsible for their renormalization.

Let us first frame the analysis of the vacuum and tadpole diagrams in preparation for section 4.1. The vacuum-to-vacuum amplitude Fig. 6 (a) takes the form of the cosmological constant term and diverges (see, e.g., in \cite{Weinberg2} and \cite{Park:2016zgt}). (The discussion here is for a flat spacetime, but the divergence will be quite generically produced for an arbitrary background.) Thus if we were dealing with a massive theory it would take a counter-term of the form of the cosmological constant of an infinite value to remove the divergence. However, the vacuum energy diagram vanishes due to an identity (eq. \rf{lid} below) in dimensional regularization. This is a rather undesirable feature of dimensional regularization when dealing with a massless theory.\footnote{For instance, the identities in \rf{vmi} and \rf{lid} often obscure cancellations between the bosonic and fermionic amplitudes in a supersymmetric field theory, making them separately vanish.} 
The diagrams responsible for the renormalization of the Newton's constant are the tadpole diagrams. As we will see in detail in section 4.1, the (would-be) shift in the Newton's constant is caused by a diagram that results from self-contraction of two fluctuation fields within the given vertex. Again, the following identity makes the regularization less suitable for the tadpole diagrams:
\bea
\int d^D k \fr1{(k^2)^\w}=0 \la{vmi}
\eea
where $\w$ is an arbitrary number. The tadpole diagram vanishes due to this: the divergence that would otherwise renormalize the Newton's constant is taken to vanish.
 For the reasons to be explained, we will introduce the shifts in the cosmological and Newton's constants through finite renormalization.

%%%%%%%%%%%%%%%%%%%%%%%%%%%%%%%%%%%%%%%%
\subsection{vacuum-to-vacuum and tadpole diagrams}
%%%%%%%%%%%%%%%%%%%%%%%%%%%%%%%%%%%%%%%%

The kinetic terms are responsible for the vacuum-to-vacuum amplitudes in the parlance of the first-layer perturbation. {We quote them here for convenience:}
\bea
&&\hspace{-.3in} 2\k^2\cL =  \sqrt{-\gt}\,\Big( -\fr12\tilde{\N}_\g h^{\a\b}\tilde{\N}^\g h_{\a\b}+\fr14 \tilde{\N}_\g h^{\a}_\a \tilde{\N}^\g h^{\b}_\b \Big)\nn\\
&&\hspace{0.1in}= -\fr12 {\pa}_\g h^{\a\b}{\pa}^\g h_{\a\b}+\fr14 {\pa}_\g h^{\a}_\a {\pa}^\g h^{\b}_\b + V_{g,I}+ V_{g,II} 
 \la{lv12q}
\eea
where
\bea
V_{g,I} &\equiv&   \Big(2\eta^{\b\b'}\tilde{\G}^{\a' \g\a}- \eta^{\a\b}\tilde{\G}^{\a' \g\b'}\Big)\pa_\g h_{\a\b}\, h_{\a'\b'}  \nn\\
V_{g,II} &\equiv& \Big[\fr12(\eta^{\a\a'}\eta^{\b\b'}\vf^{\g\g'}+\eta^{\b\b'}\eta^{\g\g'}\vf^{\a\a'}
+\eta^{\a\a'}\eta^{\g\g'}\vf^{\b\b'})\nn\\
&&-\fr14 \vf\, \eta^{\a\a'}\eta^{\b\b'}\eta^{\g\g'}-\fr12 \eta^{\g\g'}\eta^{\a'\b'}\vf^{\a\b}  \nn\\
&&+\fr14 (-\vf^{\g\g'}+\fr12 \vf \eta^{\g\g'})\eta^{\a\b}\eta^{\a'\b'}
\Big] \pa_\g h_{\a\b}\, \pa_{\g'}h_{\a'\b'}  \la{lv12}
\eea
\bea
\hspace{-.2in}{V_{g,III}} = \sqrt{-\gt}\Big( h_{\a\b}h_{\g\d}\Rt^{\a\g\b\d}-h_{\a\b}h^{\b}{}_\g \Rt^{\k\a\g}{}_{\k}
-\fr12 h^{\a\b}h_{\a\b}\Rt
\Big).  \la{gver}  
\eea
The vacuum-to-vacuum amplitudes in the first-layer perturbation can be split into two parts in the second-layer perturbation: the vacuum-to-vacuum amplitudes and the tadpoles.\footnote{As we will soon see, there are genuine tadpoles as well, i.e., tadpoles in the first-layer perturbation. As usual we evaluate them through the second-layer perturbation.} Let us consider the vacuum-to-vacuum amplitudes in the second-layer perturbation.
The vacuum energy - which leads to the cosmological constant renormalization - comes from
\bea 
\int \prod_x dh_{\k_1\k_2}\;e^{\fr{i}{\k'^2} \int  \;\Big( -\fr12\pa_\g h^{\a\b}\pa^\g h_{\a\b} \Big) }.
\eea
One obtains a constant term (see, e.g., the analysis given in \cite{Weinberg2}) whose divergent part (which will be denoted by {$A_0$} in \rf{wrml} below\footnote{Note that $A_0= - \mathscr{I}_{div}$ used, e.g., in \cite{Park:2016zgt}.}) is essentially the coefficient of the cosmological constant term. The calculation above leads to a quantum-level cosmological constant. Here is the difference between gravity and a non-gravitational theory. In a non-gravitational theory, appearance of a term absent in the classical action would potentially be a signal toward non-renormalizability.\footnote{Even in a non-gravitational theory, appearance of a {\em finite} number of new couplings is taken to be compatible with renormalizability. }
However, in a gravitational theory one has an additional leverage of a metric field redefinition, and we will ponder in section 4.2 the significance of the quantum shift  in the cosmological constant in the quantization framework that involves the metric field redefinition.

The evaluation of the vacuum-to-vacuum amplitude, whether it is from the graviton or the ghost (or matter), involves 
the following integral that is taken to vanish in dimensional regularization: 
\bea
\int d^4p \ln {p^2}=0. \la{lid}
\eea
Nevertheless, we introduce renormalization by finite renormalization for the following reasons. Although the expression above is taken to vanish in dimensional regularization, the vacuum energy expression (in particular $A_0$ in \rf{wrml}) will not, in general, vanish in other regularization methods for a curved background.
To better examine the behavior of the integral let us add a mass term $m^2$ that will be taken to $m^2\ra 0$ at the end,
\bea
\sim \int d^4p \ln { (p^2+m^2)}.
\eea
One can then take derivatives with respect to $m^2$ for its evaluation; the result takes the form of
\bea
A_f+A_0+A_1 m^2+A_2 m^4 \la{wrml}
\eea
where $A$'s are some $m$-indepedent constants; the finite piece, $A_f$, takes
\bea
A_f\sim  m^4\ln m^2. 
\eea
With the limit $m^2\ra 0$, only the term with the constant  $A_0$, which is infinite, survives, and in dimensional regularization one sets $A_0=0$. Although each term in \rf{wrml} either vanishes or is taken to zero, not introducing nonvanishing finite pieces seems unnatural (and unlikely to be consistent with the experiment): in a more general procedure of renormalization of a quantum field theory, one can always conduct finite renormalization regardless of the presence of the divergences. (As we will see in section 4.2, not only does the quantum shift  need to be introduced but also ``classical" piece of the cosmological constant.)  
Once a finite piece is introduced and the definition of the physical cosmological constant is made (say, as the coefficient of the $\int \sqrt{-\gt}$ term), the renormalized coupling will run basically due to the presence of the scale parameter $\m$.

\vspace{.1in}
Let us now consider the tadpole diagrams; the tadpole diagrams\footnote{Typically, tadpole diagrams in a non-gravitational are cancelled by a counterterm linear in the field and not considered further. More care is needed in a gravitational theory since the counter-terms take the form of the Einstein-Hilbert term. At least a priori it seems safer to view its effect as shifting the Newton's constant. The shift can be set to zero later if, for instance, the consistency of the renormalization program demands its vanishing.} are responsible for the renormalization of the Newton's constant.
For the tadpole, the rest of vertices in the kinetic term in \rf{lv12q} - which are nothing but  {$V_{g,I}$ and $V_{g,II}$} - as well as $V_{g,III}$ are relevant; the former {are} part of the vacuum-to-vacuum amplitude in the first-layer perturbation whereas the latter is associated with a genuine tadpole of the first-layer perturbation.
It turns out that {$V_{g,I},V_{g,II}$ lead to vanishing results} in dimensional regularization; we illustrate that with $V_{g,I}$,
\bea
V_{g,I}=\Big(2\eta^{\b\b'}\tilde{\G}^{\a' \g\a}- \eta^{\a\b}\tilde{\G}^{\a' \g\b'}\Big)\pa_\g h_{\a\b}\, h_{\a'\b'}.  \la{vex} 
\eea
The self-contraction of the $h_{\m\n}$'s in \rf{vex} leads to a momentum loop integral with an odd integrand, which thus vanishes. (The other terms in \rf{lv12q} vanish because the self contraction leads to the trace of $\vf_{\m\n}$.)
The vertex $V_{g,III}$ similarly leads to a vanishing result. To see this, consider contraction of the $h_{\a\b}$-fields in $V_{g,III}$. The index structures yield $R$ but the self-contraction is taken to vanish in dimensional regularization due to the identity in \rf{vmi}. Then for the renormalization of the Newton's constant the story goes similarly to the case of the cosmological constant: although the dimensional regularization does not lead to a divergence for the tadpole diagram, the shift is introduced through finite renormalization.

%%%%%%%%%%%%%%%%%%%%%%%%%%%%%%%%%%%%%%%%%%%%
\subsection{renormalization by field redefinition}
%%%%%%%%%%%%%%%%%%%%%%%%%%%%%%%%%%%%%%%%%%%%

The full one-loop renormalization procedure is in order. Many steps of the procedure below have analogous steps in the Einstein-scalar case studied in \cite{Park:2016zgt} and \cite{Park:2016vam}. Presently we put more efforts in keeping track of the finite parts, and comparison with the future experimental results is elucidated in more detail. 

Combining all the results so far, the renormalized action plus the counter-terms are given by
\bea
  &&\hspace{.3in}\int \sqrt{-g}\;(e_1+ e_2 R+ e_3 R^2+e_4 R_{\a\b}^2) \nn\\
  &&\hspace{-.3in}+\int \sqrt{-g}\Big( 
    e_5  F_{\m\k}F_\n{}^{\k}  R^{\m\n}   +e_6 F_{\a\b}F^{\a\b}R  + e_7 F_{\a\d}F_{\b\g}R^{\a \b\g\d}  \quad
    \la{totctr}
\eea    
\vspace{-.2in}
\[    \;\;
       +e_8 F_{\a\b}F_{\g\d}R^{\a \b\g\d}
  + e_{9} \nabla^\a F_{\a\k}\nabla^\b F_\b{}^\k  +e_{10}  \nabla_\l F_{\m\n}\nabla^\l F^{\m\n}  
  + e_{11}(F_{\a\b}F^{\a\b})^2 +\cdots
  \Big) 
\]
where $e_1$ is the constant previously denoted by $A_0$. More precisely, $[e_1]=A_0$ where the square bracket $[e_i]$ denotes the infinite parts of the coefficient $e_i$ calculated by employing dimensional regularization. Similarly, the would-be divergence of the tadpole diagrams will be denoted $B_0=[e_2]$. ($A_0, B_0$ are taken to vanish in dimensional regularization.) For the rest of the coefficients, one has, by collecting the results in section 3, 
\bea
&&   [e_3]=-\fr{17}{60}+\fr{23}{80}+\fr1{30},\quad    
                                         [e_4]= -\fr{7}{30}+\fr{23}{40}-\fr1{10}-\fr1{15},\quad \nn\\
&&       [e_5]=\Big(-1+\fr34\Big) \k'^2,\quad    [e_6]=\fr{\k'^2}8,\quad 
[e_7]=\fr{\k'^2}4,\quad    [e_8]=-\fr{\k'^2}4,\quad  \nn\\
&&    [e_{9}]=\fr{\k'^2}6,\quad    [e_{10}]= -\fr{\k'^2}{24} ,\quad   
[e_{11}]=\fr{3}{64}\k'^4,\quad 
\eea
where the common factor $ \fr{\G(\ve)}{(4\pi)^2}$ has been suppressed.
The finite pieces of each coefficient are determined by the $\overline{\mbox{MS}}$ scheme. 
Not all these counter-terms are independent because of the following relationships, the second of which is valid up to total derivative terms:
\bea
F_{\a\b}F_{\g\d}R^{\a \b\g\d} &=& \N_\m F_{\n\r} \N^\m F^{\n\r} +2 F_{\m\k} F_\n{}^\k R^{\m\n}-2 \N^\l F_{\l \k} \N^\s F_{\s}{}^\k \nn\\
F_{\a\d}F_{\b\g}R^{\a \b\g\d} &=& -\fr12 F_{\a\b}F_{\g\d}R^{\a \b\g\d}.
\eea
Upon substituting these into \rf{totctr}, one gets\footnote{The analysis in this section is to illustrate the renormalization procedure and is based on the computation that we have carried out in the previous sections. Some of the diagrams that we did not explicitly calculate will change the numerical values of certain coefficients.}
\bea
  &&\hspace{-.3in} \int \sqrt{-g} \; \Big[ e_1+ e_2 R+ e_3 R^2+e_4 R_{\a\b}^2   +(e_5-e_7+2e_8)  F_{\m\k}F_\n{}^{\k}  R^{\m\n}  \nn\\
  && \quad  +e_6  F_{\a\b}F^{\a\b}R 
  + (e_7-2e_8+e_{9} )  \nabla^\a F_{\a\k}\nabla^\b F_\b{}^\k  \quad      \la{totctrind}
\eea    
\vspace{-.2in}
\[    
  \;\;  +(-e_7/2+e_8+e_{10} )\nabla_\l F_{\m\n}\nabla^\l F^{\m\n}  
  + e_{11}(F_{\a\b}F^{\a\b})^2 +\cdots
  \Big].
\]
The strategy is to absorb these counter-terms by redefining the metric in the bare action. Inspection reveals that the counter-terms of the forms $\nabla_\l F_{\m\n}\nabla^\l F^{\m\n},\nabla^\a F_{\a\k}\nabla^\b F_\b{}^\k$ cannot be absorbed by a bare action that consists of the Einstein-Hilbert term and the Maxwell term: one needs the cosmological constant term as well. 
The reason is that under a metric shift $g_{\m\n}\ra g_{\m\n}+\d g_{\m\n}$, the Einstein-Hilbert part shifts according to
\bea
\sqrt{-g}\,R \ra \sqrt{-g}\,R+R\,\d g^{\m\n} \fr{\d \sqrt{-g}}{\d g_{\m\n}}+\sqrt{-g}\,\d g^{\m\n} \fr{\d R}{\d g_{\m\n}}
\eea
so the shifted part comes either with $R$ or $R_{\m\n}$, and is thus inadequate to absorb the aforementioned counter-terms; so is the shifted part from the Maxwell's action. We assume the presence of the cosmological constant in the bare action and proceed; more in the conclusion. 
Let us consider the following shifts\footnote{One may wonder about the traceless condition on the newly defined metric. The traceless condition was in order for the propagator to be well-defined. Once the effective action is obtained, one may choose a different gauge-fixing for solving the field equations in which the traceless condition may not be imposed.} \cite{tHooft:1973bhk}\cite{Park:2016zgt}, 
\[
\k\ra \k+\d\k    \quad,\quad     \L\ra \L+\d\L
\]
\bea
g_{\m\n} & \ra&  \mathscr{G}_{\m\n} \equiv l_0 g_{\m\n}+l_1  g_{\m\n}R+l_2 R_{\m\n} 
        + l_3g_{\m\n}F_{\r\s}^2 +l_4 F_{\m\k}F_\n{}^\k   \nn\\
        &&\hspace{.45in}+l_5 R F_{\m\k}F_\n{}^\k  +l_6 R_{\m\n} F_{\k_1\k_2}^2
                        + l_7 g_{\m\n} R F_{\r\s}^2 +  l_8 g_{\m\n} R^{\a\b}F_{\a\k}F_\b{}^\k  \nn\\ 
                &&\hspace{.45in}  + l_9 R_{\m}{}^\a{}_\n{}^\b F_{\a\k}F_\b{}^\k +l_{10}R  (F_{\k_1\k_2} F^{\k_1\k_2} )^2          \nn\\       
            &&  \hspace{.45in}  +l_{11}\nabla_\m F_{\k_1\k_2} \nabla_\n F^{\k_1\k_2} +l_{12}\nabla^\l F_{\l\m} \nabla^\k F_{\k\n}. 
\la{ms}
\eea
One can straightforwardly show that under these, the gravity and matter sectors shift, respectively, 
\bea
&&\hspace{-.4in} -(\fr{2}{\k^2}\L)\int \sqrt{-g} + \fr1{\k^2}\int d^4 x \sqrt{-g}\;R   \ra  -{2}\Big(\fr{\L}{\k^2}+ \fr{\d\L}{\k^2}-\fr{2\d\k \L}{{ \k^3}}+2l_0 \L \Big) \int \sqrt{-g}  \nn\\
& &\hspace{-.4in} 
+ \Big(\fr1{\k^2} -\fr{2\d \k}{\k^3} +\fr{l_0}{{ \k^2}}-\fr{\L}{{ \k^2}} (4l_1+l_2)\Big)\int \sqrt{-g}\;R
+{ \fr1{\k^2}} \int  \sqrt{-g}\Big[(l_1+\fr12 l_2)R^2-l_2 R_{\m\n}R^{\m\n}\Big] \nn\\
&&\hspace{-.2in} + { \fr1{\k^2}}\int  \sqrt{-g} \bigg[  -\L(4l_3+l_4)F_{\a\b}F^{\a\b}+\Big(l_3+l_4/2-\L[l_5+l_6+4l_7]\Big)RF_{\a\b}F^{\a\b}\nn\\
&&\hspace{-.2in}-\L(4l_8+l_9)R^{\a\b}F_{\a\k}F_\b{}^\k  -4\L l_{10}(F_{\r\s}F^{\r\s})^2  - \L l_{11}(\nabla_\m F_{\n\r})^2 -\L l_{12}(\nabla^\k F_{\k\n})^2 +\dots
\bigg]\nn\\
\eea
and\footnote{It is likely that the counter-term of the form $F_{\a\k}F_\b{}^\k F^{\a\k'}F^\b{}_{\k'}$ will appear at two-loop.}
\bea
&&\hspace{-.3in}-\fr14\int \sqrt{-g}\; F_{\m\n}^2 \ra -\fr14\int \sqrt{-g}\; F_{\m\n}^2  +\int \sqrt{-g}\;\bigg[
 -\fr{l_2}{8} RF_{\a\b}F^{\a\b} \nn\\
 &&+\fr{l_2}{2} R^{\a\b}F_{\a\k}F_\b{}^\k +\fr{l_3}{2}(F_{\r\s}F^{\r\s})^2 +\fr{l_4}{2}F_{\a\k}F_\b{}^\k F^{\a\k'}F^\b{}_{\k'}
 +\cdots\bigg]. 
\eea
Combining these two, one gets
\bea
&&\hspace{-.5in} \fr1{\k^2}\int d^4 x \sqrt{-g}\;(R-2\L)  -\fr14\int \sqrt{-g}\; F_{\m\n}^2 \ra -{2}\Big(\fr{\L}{\k^2}+ \fr{\d\L}{\k^2}-\fr{2\d\k \L}{{ \k^3}}+2l_0 \L \Big) \int \sqrt{-g}  \nn\\
& & \hspace{.2in} 
+ \Big(\fr1{\k^2} -\fr{2\d \k}{\k^3} +{ \fr1{\k^2}}l_0-{ \fr1{\k^2}}\L (4l_1+l_2)\Big)\int \sqrt{-g}\;R  -\fr14\int \sqrt{-g}\; F_{\m\n}^2
 \nn\\
&&\hspace{-.2in} + { \fr1{\k^2}}\int  \sqrt{-g}\Big[(l_1+\fr12 l_2)R^2-l_2 R_{\m\n}R^{\m\n}\Big]+{ \fr1{\k^2}} \int  \sqrt{-g} \bigg[  -\L(4l_3+l_4)F_{\a\b}F^{\a\b}
\nn\\
&&\hspace{-.2in} +\Big(l_3+\fr{l_4}2-\L[l_5+l_6+4l_7] -\fr{l_2}{8} { \k^2}\Big)RF_{\a\b}F^{\a\b}+\Big({ {\k^2}}\fr{l_2}{2}-\L[4l_8+l_9]\Big)R^{\a\b}F_{\a\k}F_\b{}^\k  \nn\\
&& ({ \k^2} l_3/2 -4\L l_{10})(F_{\r\s}F^{\r\s})^2 -l_{11}(\nabla_\m F_{\n\r})^2 -l_{12}(\nabla^\k F_{\k\n})^2 +\dots
\bigg]. 
\eea
Not all of the terms in the expansion {have been} explicitly recorded: additional diagrams such as 3-pt amplitudes should be considered to account for some of them.

Let us consider the first several coefficients of the shifted action and compare them with those of \rf{totctrind}. We start with the cosmological constant term and  Einstein-Hilbert term. Their counter-terms can be absorbed by setting 
\bea
-\fr{2}{\k^2}\Big(\d\L-\fr{2\d\k  \L}{{ \k}}\Big) =  A_0   \la{dL}
\eea
and 
\bea
-\fr{2}{{ \k^3}}\d\k +\fr1{\k^2}l_0 - \fr{\L}{\k^2}  (4l_1+l_2)
= B_0
\la{dk}
\eea
respectively. We assume that the constants $A_0,B_0$ now contain the non-vanishing finite pieces introduced by the aforementioned finite renormalization. Eq. \rf{dk} determines the infinite part of $\d \k$
\bea
 \d\k =\fr{\k}{2}l_0 - \fr{\k\L}{2}  (4l_1+l_2)  - \fr{\k^3}{2}B_0.
\eea
 $\d \L$ is determined once this result is substituted into \rf{dL}: 
 \bea 
 \d \L=l_0 \L-\L^2 (4l_1+l_2)-\fr{\k^2}2 A_0.
 \eea
 The counter-terms of the forms $R^2,R_{\m\n}^2$ can be absorbed by setting
\bea
 l_1+\fr12 l_2=e_3
	\quad,\quad - l_2= e_4 \nn\\
\eea
which yields
\bea
l_1= e_3+\fr12e_4 \quad,\quad  l_2= -e_4.
\eea	
Inspection of the coefficients of {$F_{\a\b}^2$} implies
\bea
4l_3+l_4=\cO(\k^4).
\eea
The coefficients of $RF_{\a\b}F^{\a\b}$, $R^{\a\b}F_{\a\k}F_\b{}^\k$ should match with the corresponding coefficients of the counter-term action: 
\[
 l_3+\fr{l_4}2-\L  (l_5+l_6+4l_7)-\fr{l_2}8 { \k^2}=e_6\;,\; \fr{{ \k^2}}2 l_2-\L  (4l_{8}+l_{9}) =e_5-e_7+2e_8. \nn\\
\]
These constraints are to be combined with those coming from the higher order counter-terms.

%%%%%%%%%%%%%%%%%%%%%%%
%%%%%%%%%%%%%%%%%%%%%%%
\subsection{beta function analysis}

In section 4.1 we have carried out the analysis without including the cosmological constant.
As reviewed therein, dimensional regularization has a technical subtlety: the flat propagator yields vanishing results for the vacuum and tadpole diagrams. For this reason, the shifts in the coupling constants were introduced through finite renormalization. Through the analysis in section 4.2, it has been shown that the cosmological constant is generically generated by the loop effects and the renormalizability requires its presence in the bare action. 
This status of the matter suggests the possibility of carrying out an alternative renormalization procedure in which the cosmological constant is included in the starting renormalized action. Once the cosmological constant is included and expanded around the fluctuation metric, the third and higher order terms can be treated as a source for additional vertices. The  second-order fluctuation term, however, can be treated as the ``graviton mass" term.\footnote{As a matter of fact,  we have become aware, after publication of the present work, of the work \cite{Toms:2008dq} in which it was noted that the presence of a cosmological constant led to running of the matter coupling constant that was absent when the cosmological constant was not present. It was observed that the cosmological constant acts like mass terms for photon and graviton, a fact independently noted in \cite{Park:2016zgt} for the graviton.} With this arrangement, the vacuum-to-vacuum and tadpole diagrams yield non-vanishing results. In this subsection we carry out the beta function analysis of the vector coupling constant and illustrate this alternative procedure. 

An analysis of renormalization of the vector coupling was previously carried out in \cite{Huggins:1987zw} by employing the setup of \cite{Vilkovisky:1984st}. It was observed that the presence of the cosmological constant generates the formal mass terms for the photon and graviton. Our beta function calculation below yields the same result as that of \cite{Toms:2008dq}.

In the present context, treating a cosmological constant-type term as the graviton mass term was considered in \cite{Park:2016zgt} for an Einstein-scalar theory with a Higgs-type potential. Let us similarly treat the quadratic part of the cosmological constant term as a formal mass term for the graviton: 
\be
m^2 =-2\L.
\ee
For the detailed analysis of the beta function, it is convenient, and common, to introduce a scale parameter $\m$ by making the following scalings
%\footnote{One can similarly rescale the vector coupling constant $e$; for the one-loop beta function of the vector coupling $e$ only the graviton loop and thus rescaling of $\k$ is relevant.}
:
\bea
\k^2\ra \m^{2\ve}\k^2\quad,\quad  e^2\ra \m^{2\ve}e^2\quad,\quad \L\ra \m^{-2\ve}\L.
\eea
With this, eq. \rf{EM} takes
\bea
S=\int \sqrt{-\gh}\;\Big(\fr1{\k^2 \m^{2\ve}}(\Rh-2\L \m^{-2\ve} )-\fr1{4e^2 \m^{2\ve}} \Fh_{\m\n}^2 \Big)
\eea
and one gets, for the kinetic part of the gravity sector, 
\bea
\cL_{kin} = \fr1{2\k^2 \m^{2\ve} }\sqrt{-\gt}\Big[ -\fr12{\tilde{\N}}_\g h^{\a\b}{\tilde{\N}}^\g h_{\a\b}
%+\fr14 {\tilde{\N}}_\g h^{\a}_\a {\tilde{\N}}^\g h^{\b}_\b  
 -\fr12 (-2\L)h_{\m\n}h^{\m\n})  \Big].
\eea
Treating the cosmological constant-containing term as the ``mass" term modifies \rf{thickprop} to
\bea
\tilde{\D}(X_1-X_2)=\int \fr{d^4L}{(2\pi)^4}\fr{e^{iL_{\underline{\d}} (X_1-X_2)^{\underline{\d}}}}{i (L_{\underline{\a}} L_{\underline{\b}} \h^{\underline{\a}\underline{\b}}-2\L)}.
\la{fspm}
\eea
The correlator relevant for the vector coupling renormalization is
\bea
 &&
 -\fr{i}{4\m^{2\ve}e^2}\int \sqrt{-\gt} \Ft_{\m\r}\Ft_{\n\s}      \Big<
\gt^{\m\n}h^{\r\k}h_\k^{\s}  +\gt^{\r\s}h^{\m\k}h_\k^{\n} -\fr12 \gt^{\m\n}hh^{\r\s}  -\fr12 \gt^{\r\s}hh^{\m\n}\nn\\
&&\hspace{1.2in} +h^{\m\n}h^{\r\s}  +\fr18 \gt^{\m\n} \gt^{\r\s}(h^2-2h_{\k_1\k_2}h^{\k_1\k_2} )
\Big>.    \la{vcbf}  
\eea
Carrying out the self-contractions of the fluctuation fields one gets  
\bea
 %\hspace{-.3in}
&=& -\fr{3i\m^{2\ve}\k'^2}{8\m^{2\ve}e^2}\int \sqrt{-\gt} \;\Ft_{\m\r}\Ft^{\m\r}      \int \fr{d^4L}{(2\pi)^4}\fr{1}{i ( L^2-2\L \m^{-2\e})} \nn\\
&=& -\fr{3\m^{2\ve}\k'^2}{8\m^{2\ve}e^2} \fr{\G(\ve) (2\L  \m^{-2\ve})}{(4\pi)^2}  \int \sqrt{-\gt} \;\Ft_{\m\r}\Ft^{\m\r}   \nn\\
&\simeq& \fr{3}{32\pi^2(D-4)} \fr{ \k^2 \L  }{\m^{2\ve}e^2}   \int \sqrt{-\gt} \;\Ft_{\m\r}\Ft^{\m\r}  
\eea
 where the second equality has been obtained by performing the momentum integration after a Wick rotation. The third equality is obtained by keeping only the pole term in expansion of $\G(\ve)$. The result above implies that the one-loop-corrected vector coupling $e_1$ is given by
 \bea
 e_1=e \m^{\ve}\Big( 1+ \fr{3}{8\pi^2(D-4)} {\k^2 \L  }  \Big)^{\fr12}\simeq e \m^{\ve}\Big( 1+ \fr{3}{16\pi^2(D-4)} {\k^2 \L  }  \Big).
 \eea 
 From this it follows that
 \be
\m \fr{\pa  e_1}{\pa \m}=\ve e \m^\ve -\fr{3\m^\ve }{32\pi^2} \k^2 \L\, e.
 \ee
Taking $\ve\ra 0$, one gets the following beta function: 
 \be
\b(e)= -\fr{3 }{32\pi^2} \k^2 \L e.  \la{bfc}
 \ee 
 This is the same result as that obtained in \cite{Toms:2008dq} because $\k^2$ in \cite{Toms:2008dq} is twice $\k^2$ here.\footnote{The result in \cite{Toms:2008dq} was obtained by employing the standard one-loop determinant formula. Although, strictly speaking, the formula makes sense only when the traceless part of the fluctuation field is taken out, once the formula is used it doesn't matter how it was obtained. To rephrase, the result in \cite{Toms:2008dq} was obtained, so to speak, by bypassing the traceful propagator. We believe that this is why the present beta function result - obtained by employing the traceless propagator - agrees with that therein obtained. Incidentally there is a coincidence: it turns out that the result \rf{vcbf} remains the same even if one employs the traceful propagator.}

In an earlier related work by Robinson and Wilczek \cite{Robinson:2005fj} a potentially interesting idea of a shift in the unification scale due to the quantum gravity contributions to the gauge coupling was proposed. The idea was further debated \cite{Pietrykowski:2006xy}\cite{Toms:2010vy}\cite{Ellis:2011} and the possibility of QED asymptotic freedom was proposed \cite{Toms:2010vy}. Before we get to the implications of the present work for these proposals let us note several technical differences between the beta function analysis of \cite{Robinson:2005fj} and the one above.\footnote{The system considered in \cite{Robinson:2005fj} is a non-Abelian gauge theory coupled to gravity. For this reason there are more graphs to be considered. For instance the graph in Fig. 1 there - which requires the cubic gauge coupling - does not arise in the present case. If one considers a Higgs type scalar field as well, then the gauge field becomes massive and the graph should produce an additional contribution analogous to the one obtained in \cite{Toms:2010vy}. For that matter, it will also be interesting to investigate whether and how finite temperature effects lead to a term analogous to the one obtained in \cite{Toms:2010vy}.} One difference is the use of traceless/traceful propagator. Another difference is the regularization scheme: momentum cutoff in \cite{Robinson:2005fj} vs. dimensional regularization in the present work. For a massless theory certain graphs vanish in dimensional regularization due to the identities such as \rf{vmi} and \rf{lid} whereas they do not in momentum cutoff. Still another difference is the gauge-fixing.   

In the analysis \cite{Toms:2010vy} performed with heat kernel method and momentum cutoff, an additional term to the beta function was obtained. The additional term has the same form as the existing term with $\L$ replaced by a characteristic energy scale square. It then led to potential asymptotic freedom in QED. This proposal as well as that of \cite{Robinson:2005fj} were subsequently debated in \cite{Pietrykowski:2006xy} and \cite{Ellis:2011}. In particular, it was suggested in \cite{Pietrykowski:2006xy} such an additional term will be absent in the de Donder gauge. This suggests the possibility that the potential gauge choice-dependence may be responsible for the difference. Given the $\xi$-dependence discussed in section 3.2, we believe that it is a reasonable possibility.

%%%%%%%%%%%%%%%%%%%%%%%%%%%%%%%%%%%%%%%%
\subsection{scattering predictability and more on 1PI action}
%%%%%%%%%%%%%%%%%%%%%%%%%%%%%%%%%%%%%%%%

It may be useful to recall the case of a non-gravitational theory before considering the gravitational case.
Suppose one performed the loop computations and found new vertices required to remove the divergences. In the standard procedure of renormalization, those vertices will be included with the corresponding arbitrary coupling constants in the bare action. 
If the number of the new vertices is finite, the theory is called renormalizable and one proceeds to obtain the 1PI effective action. If the number is infinite, the theory is declared to be unrenormalizable; the infinite number of coupling constants lead to loss of the predictive power of the theory. 
 
 In a gravity theory there are an infinite number of counter-vertices, some of which we have seen in section 3. 
The idea of the field redefinition is that one starts with the bare action of the same form as the classical action with possible addition of the cosmological constant term. The metric in the bare action is the field-redefined one, $\mathscr{G}_{\m\n}$, in \rf{ms}. 
The crucial point is that all of the coupling constants associated with the higher-order counter-vertices are absorbed into the {\em redefined} metric $\mathscr{G}_{\m\n}$, thus unobservable \cite{tHooft:1973bhk}.
 The predictability of the theory for scattering amplitudes then follows.
 Let us paraphrase. The divergences arising from the loop diagrams can be removed by the counter-vertices present on the right-hand side of the definition of $\mathscr{G}_{\m\n}$ in \rf{ms}. In other words, with the counter-terms added, the renormalized action now contains all of the new coupling constants. However, the counter-terms with those coupling constants can be combined into Einstein-Hilbert form in terms of the redefined metric $\mathscr{G}_{\m\n}$. This means that the bare action has two coupling constants, the cosmological and Newton's, in terms of the new metric, and therefore the theory is predictive. 

More specifically, the theory should become predictive by following the usual ``routines": suppose the experimental values of the cosmological and Newton's constants are known accurately to the extent that we may discern the quantum corrections. One can find the values of the renormalized cosmological and Newton's constants by imposing certain renormalization conditions. Once the renormalized constants are determined in terms of physical constants one can proceed to compute, for instance, various scattering amplitudes and make predictions on the corresponding experimental outcomes.

The fact that the infinite number of the coupling constants are absorbed by the metric field redefinition must not be taken as to mean that the quantum effects are immaterial. There will be all those vertices present with fixed finite values of the coefficients in the effective action.\footnote{In general a full effective action is a highly complicated object even containing nonlocal terms (that may be important for the black hole physics \cite{Mukhanov:1994ax}). Here we focus on the starting renormalized action with the added vertices with the fixed finite coefficients.} At this point one can consider yet another field redefinition in conjunction with the quantum deformation of the geometry that plays an important role in the context of black hole information \cite{Park:2017dib}\cite{Park:2017wiw}\cite{Nurmagambetov:2018het}.

%%%%%%%%%%%%%%%%%%%%%%%%%%%%%%%%%%%%%%%%%%%%
%%%%%%%%%%%%%%%%%%%%%%%%%%%%%%%%%%%%%%%%%%%%
\section{Conclusion}
%%%%%%%%%%%%%%%%%%%%%%%%%%%%%%%%%%%%%%%%%%%%
%%%%%%%%%%%%%%%%%%%%%%%%%%%%%%%%%%%%%%%%%%%%

In this work, we have extended the one-loop renormalization of an Einstein-scalar system to an Einstein-Maxwell system. 
As in the previous works the amplitude calculations have been carried out with the two layers of perturbation in the refined background field method.  
Since the Maxwell part itself is a gauge system, the extension involves overcoming several additional hurdles. 
The direct Feynman diagrammatic computation has lead to a gauge choice-dependent 1PI effective action: the  effective action is covariant only up to the metric gauge-fixing. The origin of the dependence was found in the limitation of the background field method. The proper interpretation of the particular gauge-choice independence that we have focused on in this work is such that the effective action is made covariant by removing all of the terms containing the gauge-fixing condition. At the same time, the action is to be supplemented by the gauge-fixing condition in the usual manner. We also discussed how the present formalism allows one to avoid or re-interpret the well-known gauge-choice dependence. We have taken one step further compared to our previous works: with the fixed renormalization scheme chosen, we have enumerated the quantum corrections of various physical quantities such as the cosmological and Newton's constants. The role of the finite renormalization is important. We have seen that the metric field redefinition a la 't Hooft brings predictive power to the theory.

There are several highlights worth recapitulating. 
Firstly, note that one ends up taking three different measures to ensure the covariance: removal of the trace part of the metric, employment of the refined BFM, and enforcement of the strong form of the gauge-fixing.
Secondly, the cosmological constant has several special features in the context of renormalization. It is the leading term in the derivative expansion and generically generated regardless of the background under consideration. Even if one's starting action does not include the cosmological constant term, the renormalizability dictates its presence in the bare action (and thus in the effective action). 
Thirdly, the renormalizability requires a metric field definition. The existence of such a field redefinition should not be a coincidence but must be a reflection of the quantum deformation of the geometry. The freedom of such a field redefinition is powerful and distinguishes gravity from non-gravitational unrenormalizable theories.

We have seen that the renormalizability requires the presence of the cosmological term in the bare action. 
One may take this as a rationale of the presence of a renormalized cosmological constant in the starting renormalized action. In fact, this seems to suggest a future direction that stands out. In the main body we carried out the analysis without including the cosmological constant since we were interested in a flat background. The fact that the cosmological constant is generically generated and required for the renormalizability seems to suggest the possibility that it should be included in the starting renormalized action. This would imply that one should consider the propagator associated with a de Sitter (or anti-de Sitter) background, although the flat spacetime analysis can still be employed for the divergence analysis. Once the cosmological constant is included and expanded around the fluctuation metric, it can be treated as a source for additional vertices.\footnote{Alternatively,  it can be treated as the ``graviton mass" term and with this the graviton propagator becomes a massive one as we discussed in section 4.3. This may appear too contrived but the massive propagator makes it unnecessary to introduce the finite renormalization. An objection may be raised that the spacetime is no longer flat in the presence of the cosmological constant term. This is an issue worth further exploring. The bottom line is that the flat space analysis catches the divergences. Also, in a more realistic setup including a Higgs-type scalar, mixing between the physical and unphysical states \cite{Pius:2014iaa} is expected.
}

It appears that there are several variant renormalization procedures depending on, e.g., whether or not to include the cosmological constant in the starting renormalized action. As a matter of fact, there is an intriguing possibility when choosing a renormalization scheme. Although the flat space analysis catches the divergent parts of the proper curved space analysis, the finite parts require, in general, the due curved space propagators. One may choose the renormalization scheme such that the finite parts become the same as the corresponding flat analysis. It will be interesting to see whether or not the renormalization procedure could be consistently conducted with such a special scheme. If it could be and yields the same results as the curved space analysis, the flat space analysis will serve as a highly convenient alternative to the proper curved space analysis, and that would imply, in a certain sense, the background independence of the whole framework. Not unrelated, it will be interesting to see whether the aforementioned freedom in the renormalization scheme could be used to absorb the the gauge-choice dependence of the coefficients appearing in the action \rf{totctr}. (A recent review on such gauge-choice dependence can be found in \cite{Lavrov:2012xz}.)

Another direction is the two-loop extension of the results of the present work.
As stated in the introduction, the renormalization procedure in this paper is entirely within the standard framework and in particular the reduction of the physical states did not play a role (other than its role in gauge-choice independence and providing assurance that the present procedure can, in principle, be extended to two- and higher- loops). Although the direct two-loop analysis is expected to be much harder, the difficulty will be of technical character and associated with computing the Feynman diagrams themselves. One may turn to the approach where the counter-terms are determined by dimensional analysis and covariance \cite{Goroff:1985th}. Once the counter-terms are obtained one way or another, it should be possible, with reasonable effort, to extend the field redefinition-utilized renormalization to two-loop. The reduction to the physical sector \cite{Park:2014tia,Park:2015ybl} - which should be performed after offshell computation - is expected to play a role at two- and higher- loops.

%\vspace{.4in} 

%\ni {\bf Note added:}

\newpage
\appendix

\newpage
%%%%%%%%%%%%%%%%%%%%%%%%%%%%%%%%%%%%%%%%%%%%%%%%%%%%%%%%%%%%%%%%

\end{document}